\documentclass[reprint,showpacs,preprintnumbers,nofootinbib,amsmath,amssymb,aps,prd,floatfix]{revtex4-1}
\pdfoutput=1
\usepackage{graphicx}
\usepackage{dcolumn}
\usepackage{bm}
\usepackage{subfigure}
\usepackage{url}
\usepackage[hyperindex]{hyperref}
\usepackage{color}
\usepackage[ddmmyy,24hr]{datetime}
\newcommand{\nua}[1]{\ensuremath{\rlap{\kern-2.5pt\ensuremath{\overset{\scriptscriptstyle(-)}{\phantom{\nu}}}}{\ensuremath{{\nu}_{#1}}}}}
\newcommand{\vet}[1]{\ensuremath{\hskip-1pt\vec{\hskip1pt#1}}}
\begin{document}

\preprint{CERN-TH-2016-025}
\preprint{DESY 16-023}

\title{Assessing the role of nuclear effects in the interpretation of the MiniBooNE low-energy anomaly}

\author{M. Ericson}
\affiliation{Universit\'e de Lyon, Univ. Lyon 1, CNRS/IN2P3, IPN Lyon, F-69622 Villeurbanne Cedex, France}
\affiliation{Physics Department, Theory Unit, CERN, CH-1211 Geneva, Switzerland}

\author{M. V. Garzelli}
\affiliation{II Institute for Theoretical Physics, Hamburg University, Luruper Chaussee 149, D--22761 Hamburg, Germany}

\author{C. Giunti}
\affiliation{INFN, Sezione di Torino, Via P. Giuria 1, I--10125 Torino, Italy}

\author{M. Martini}
\affiliation{ESNT, CEA-Saclay, IRFU, Service de Physique Nucl\'eaire, F-91191 Gif-sur-Yvette Cedex, France}

\date{\dayofweekname{\day}{\month}{\year} \ddmmyydate\today, \currenttime}

\begin{abstract}
We study the impact of the effect of multinucleon interactions
in the reconstruction of the neutrino energy
on the fit of the MiniBooNE data in terms of neutrino oscillations.
We obtain some improvement of the fit
of the MiniBooNE low-energy excess in the framework of two-neutrino oscillations
and a shift of the allowed region
in the
$\sin^2 2\vartheta$--$\Delta{m}^2$
plane towards
smaller values of
$\sin^2 2\vartheta$
and
larger values of
$\Delta{m}^2$.
However
this effect is not enough to
solve the problem of the
appearance-disappearance tension
in the global fit
of short-baseline neutrino oscillation data.
\end{abstract}


\maketitle

\section{Introduction}
\label{intro}

Neutrino masses and mixing
are well-established by the observations of neutrino oscillations in
solar, atmospheric and long-baseline
neutrino oscillation experiments,
which are well-accommodated in the standard framework of
three-neutrino mixing,
where the three active neutrinos
$\nu_{e}$,
$\nu_{\mu}$,
$\nu_{\tau}$
are superpositions of three massive neutrinos
$\nu_1$,
$\nu_2$,
$\nu_3$
with respective masses
$m_1$,
$m_2$,
$m_3$
(see Refs.\cite{Giunti:2007ry,GonzalezGarcia:2007ib,Bellini:2013wra}).
In this framework there are two independent squared-mass differences:
the small solar
$\Delta{m}^2_{\text{SOL}} \simeq 7.5 \times 10^{-5} \, \text{eV}^2$
and the larger atmospheric
$\Delta{m}^2_{\text{ATM}} \simeq 2.3 \times 10^{-3} \, \text{eV}^2$,
which can be interpreted as
$
\Delta{m}^2_{\text{SOL}}
=
\Delta{m}^2_{21}
$
and
$
\Delta{m}^2_{\text{ATM}}
=
|\Delta{m}^2_{31}|
\simeq
|\Delta{m}^2_{32}|
$,
with
$\Delta{m}^2_{kj}=m_k^2-m_j^2$.

However,
the
reactor \cite{Mention:2011rk,Mueller:2011nm,Huber:2011wv},
Gallium \cite{Abdurashitov:2005tb,Laveder:2007zz,Giunti:2006bj,Giunti:2010zu,Giunti:2012tn}
and
LSND \cite{Aguilar:2001ty}
anomalies indicate that the neutrino mixing framework
may need an extension in order to accommodate
short-baseline oscillations\footnote{
Short-baseline neutrino oscillation experiments
are characterized by a ratio
$L/E_{\nu} \lesssim 10 \, \text{m} \, \text{MeV}^{-1}$,
where
$L$ is the source-detector distance
and
$E_{\nu}$ is the neutrino energy.
Since the oscillations generated by a
squared-mass difference $\Delta{m}^2$ is observable for
$\Delta{m}^2 L / 4 E_{\nu} \gtrsim 1$,
short-baseline neutrino oscillation experiments are sensitive to
$\Delta{m}^2 \gtrsim 10^{-1} \, \text{eV}^2$.
On the other hand,
long-baseline neutrino oscillation experiments
are characterized by a ratio
$L/E_{\nu} \gtrsim 100 \, \text{m} \, \text{MeV}^{-1}$
which makes them sensitive to
$\Delta{m}^2 \lesssim 10^{-2} \, \text{eV}^2$.
}
due to at least one additional squared-mass difference,
$\Delta{m}^2_{\text{SBL}} \sim 1 \, \text{eV}^2$,
which is much larger than $\Delta{m}^2_{\text{ATM}}$
(see the reviews in Refs.~\cite{Bilenky:1998dt,Strumia:2006db,GonzalezGarcia:2007ib,Abazajian:2012ys,Conrad:2012qt,Palazzo:2013me,Gariazzo:2015rra}).
The reactor antineutrino anomaly
\cite{Mention:2011rk}
is a deficit of the rate of $\bar\nu_{e}$-induced events observed in
short-baseline reactor neutrino experiments
in comparison with that expected from the calculation of
the reactor neutrino fluxes
\cite{Mueller:2011nm,Huber:2011wv}.
The Gallium neutrino anomaly
\cite{Abdurashitov:2005tb,Laveder:2007zz,Giunti:2006bj,Giunti:2010zu,Giunti:2012tn}
is a shortage of $\nu_{e}$-induced events
measured at an average distance of about 1 m in the
Gallium radioactive source experiments
GALLEX
\cite{Kaether:2010ag}
and
SAGE
\cite{Abdurashitov:2009tn}
with respect to the rate of $\nu_{e}$-induced events
expected from the well-measured activity of the radioactive source.
The
LSND anomaly
is the observation of short-baseline
$\bar\nu_{\mu}\to\bar\nu_{e}$
transitions
\cite{Athanassopoulos:1995iw,Aguilar:2001ty}.

The additional squared-mass difference
required to explain these anomalies with neutrino oscillations
necessitates the existence of at least an additional massive neutrino
at the eV scale.
Since from the LEP measurement
of the invisible width of the $Z$ boson \cite{ALEPH:2005ab}
we know that there are only three active neutrinos,
in the flavor basis the additional massive neutrinos correspond to
sterile neutrinos
\cite{Pontecorvo:1968fh},
which do not have standard weak interactions.

Sterile neutrinos are singlets of the Standard Model gauge symmetries
which can couple to the active neutrinos through the Lagrangian mass term.
In practice there are bounds on the active-sterile mixing,
but there is no bound on the number of sterile neutrinos and on their mass scales.
Therefore the existence of sterile neutrinos
is investigated at different mass scales\footnote{
For example:
very light sterile neutrinos at a mass scale smaller than 0.1 eV, which could affect the oscillations of solar
\cite{deHolanda:2003tx,deHolanda:2010am,Das:2009kw}
and reactor
\cite{Kang:2013zma,Bakhti:2013ora,Palazzo:2013bsa,Esmaili:2013yea,Girardi:2014gna,Girardi:2014wea,DiIura:2014csa,An:2014bik}
neutrinos;
sterile neutrinos at the keV scale,
which could constitute warm dark matter according to the Neutrino Minimal Standard Model ($\nu$MSM)
\cite{Asaka:2005an}
(see the reviews in Refs.~\cite{Drewes:2013gca,Boyarsky:2012rt});
sterile neutrinos at the MeV scale
\cite{Kusenko:2004qc,Gelmini:2008fq,Masip:2012ke,Ishida:2014wqa};
sterile neutrinos at the electroweak scale
\cite{Antusch:2015mia,Deppisch:2015qwa}
or above it
\cite{Antusch:2014woa,Deppisch:2015qwa},
whose effects may be seen at LHC and other high-energy colliders.
}.
In this paper we consider
the simplest 3+1 extension of three-neutrino mixing
in which
the three standard massive neutrinos
$\nu_1$,
$\nu_2$,
$\nu_3$
are assumed to have masses much smaller than 1 eV
and there is an additional neutrino $\nu_{4}$
with mass $m_{4} \sim 1 \, \text{eV}$.
In this framework,
the squared-mass difference
$\Delta{m}^2_{\text{SBL}} = \Delta{m}^2_{41} \sim 1 \, \text{eV}^2$
can generate short-baseline oscillations
which explain the above-mentioned anomalies
(see Ref.~\cite{Gariazzo:2015rra}).
In the flavor basis,
besides the three active neutrinos
$\nu_{e}$,
$\nu_{\mu}$,
$\nu_{\tau}$,
there is a sterile neutrino
$\nu_{s}$
which has a large mixing with $\nu_{4}$
and small mixings with
$\nu_1$,
$\nu_2$,
$\nu_3$.
This implies that the elements of
the $4\times4$ unitary mixing matrix $U$
must be such that
$|U_{sk}| \ll 1$
for $k=1,2,3$
and
$|U_{\alpha 4}| \ll 1$
for $\alpha=e,\mu,\tau$.

In this paper we consider
the results of the MiniBooNE experiment
\cite{AguilarArevalo:2008rc,Aguilar-Arevalo:2013pmq},
which has been done to check the indication in favor of
short-baseline neutrino oscillations given by the LSND anomaly\footnote{
The reactor and Gallium anomalies will be checked in a few years
by several reactor and radioactive source experiments
(see Refs.~\cite{Gariazzo:2015rra,Giunti:2015wnd}).
}.
The MiniBooNE experiment
operated first in neutrino mode,
searching for $\nu_{\mu}\to\nu_{e}$ transitions.
The results ``showed no evidence of an excess of
electron-like events for neutrino energies above 475 MeV''
\cite{AguilarArevalo:2008rc},
which cover the same $L/E_{\nu}$ range of the LSND experiment.
On the other hand,
the data showed
``unexplained electron-like events in the reconstructed neutrino energy range from 200 to 475 MeV''
\cite{AguilarArevalo:2008rc}.
This is a sizable excess of $\nu_{e}$-like events in the three energy bins below
475 MeV which has been called the \emph{MiniBooNE low-energy anomaly}.

The second part of the MiniBooNE experiment was operated in antineutrino mode,
searching for $\bar\nu_{\mu}\to\bar\nu_{e}$ transitions.
The final results
\cite{Aguilar-Arevalo:2013pmq}
showed a small excess of
$\bar\nu_{e}$-like events over the background
for reconstructed neutrino energies above 475 MeV
and a sizable excess of $\bar\nu_{e}$-like events in the three energy bins below
475 MeV
which is compatible with the low-energy anomaly in neutrino mode.

The authors of Refs.~\cite{Martini:2012fa,Martini:2012uc}
suggested that at least part of the MiniBooNE low-energy excess could be due to
events which have a larger neutrino energy and are interpreted as low-energy events
because the reconstruction of the neutrino energy
from the measured electron energy and scattering angle
did not take into account multinucleon interactions in the neutrino-nucleus charged-current interactions.

The multinucleon emission channel has attracted much attention in the last years.
The inclusion of this channel in the quasielastic cross section
was suggested \cite{Martini:2009uj,Martini:2010ex} as
the possible explanation of the MiniBooNE CCQE total cross section on carbon \cite{AguilarArevalo:2010zc}, observed to be too
large with respect to theoretical predictions employing the standard value of the axial mass. The MiniBooNE experiment, as well as other experiments involving Cherenkov detectors, defines as a charged current quasielastic-like event one in which only a final charged lepton is detected. The ejection of
a single nucleon (a genuine quasielastic event) is only one possibility, and one must consider as well events involving for instance a correlated nucleon pair from which the partner nucleon is also ejected, as discussed first in Refs.~\cite{Marteau:1999jp}.
The inclusion in the quasielastic cross section of
events in which several nucleons are ejected ($np$-$nh$ excitations), leads to an increase over the genuine quasielastic value. The authors of Refs.~\cite{Martini:2009uj,Martini:2010ex} argued that this is the likely explanation of the MiniBooNE data, showing that their evaluation can account for the excess in the cross section without any modification of the axial mass.
This suggestion triggered a new interest of the neutrino scattering and oscillation communities for the multinucleon emission channel.
Beyond the first MiniBooNE data \cite{AguilarArevalo:2010zc},
the appearance of new measurements of charged current quasielastic-like cross sections
\cite{AguilarArevalo:2013hm,Fields:2013zhk,Fiorentini:2013ezn,Wolcott:2015hda,Abe:2016tmq},
of analyses of the hadronic final states
\cite{Acciarri:2014gev,Walton:2014esl,Abe:2015oar}
and of the vertex and recoil energies deposited in the detector
\cite{Fields:2013zhk,Fiorentini:2013ezn,Rodrigues:2015hik}
is leading to a mounting experimental evidence of the multinucleon effects in neutrino-nucleus scattering.
Several theoretical works \cite{Martini:2009uj,Martini:2010ex,Amaro:2010sd,Nieves:2011pp,Bodek:2011ps,Martini:2011wp,Nieves:2011yp,Amaro:2011aa,Lalakulich:2012ac,Nieves:2013fr,Martini:2013sha,Gran:2013kda,Martini:2014dqa,Megias:2014qva,Ericson:2015cva,Ivanov:2015aya,Martini:2016eec}
have analyzed the role of the multinucleon excitations in the evaluation of the neutrino-nucleus cross sections at MiniBooNE, T2K and MINERvA energies.
Originally, this channel was not included in the Monte Carlo generators used for the analyses of the neutrino cross sections and oscillations experiments. Today there is an effort to include this $np$-$nh$ channel in several event generators~\cite{Sobczyk:2012ms,Katori:2013eoa,Schwehr:2016pvn,Wilkinson:2016wmz}.
As was discussed in Refs.~\cite{Martini:2012fa,Martini:2012uc,Nieves:2012yz,Lalakulich:2012hs},
the influence of the multinucleon channel also manifests in the problem of the neutrino energy reconstruction \cite{Leitner:2010kp,Martini:2012fa,Martini:2012uc,Nieves:2012yz,Lalakulich:2012hs,Mosel:2013fxa,Ankowski:2014yfa,Ankowski:2015jya,Coloma:2013rqa,Coloma:2013tba}.
The authors of Refs.~\cite{Martini:2012fa,Martini:2012uc} also showed how it affects the analysis of neutrino oscillation experiments.
Applied to the MiniBooNE data this is the object of the present work.

In this paper we study the impact of
the multinucleon interactions in neutrino-nucleus charged-current scattering
on the fit of the electron appearance MiniBooNE data in terms of neutrino oscillations.
The MiniBooNE collaboration discussed briefly
an approximate implementation of the multinucleon interactions
in Ref.~\cite{Aguilar-Arevalo:2013pmq}.
They showed that
the change of the minimum $\chi^2$ value
is small and they concluded that
multinucleon interactions do not significantly change the oscillation fit.
However,
the change of the minimum $\chi^2$ value
due to multinucleon interactions
obtained by the MiniBooNE collaboration is a small increase,
whereas we expect a decrease
from a better fit of the low-energy excess.
In this paper we examine this problem and we study quantitatively
to which extent the results of the oscillation fit of the MiniBooNE data
is affected by the multinucleon interactions.

The plan of the paper is as follows.
In Section~\ref{method}
we describe the method that we adopted in order to take into account
the multinucleon contribution to the neutrino-nucleus charged-current interactions
in the analysis of MiniBooNE data.
In Section~\ref{two} we present the results of the fit of MiniBooNE data
taking into account multinucleon interactions
in the simplest framework of two-neutrino mixing.
In Section~\ref{3+1} we discuss the implications
of the multinucleon interactions in the analysis of MiniBooNE data
for the global fit of short-baseline neutrino oscillation data
in the framework of 3+1 neutrino mixing.
Finally,
in Section~\ref{conclusions}
we present our conclusions.

\section{Method}
\label{method}

In principle, the multinucleon interactions should be included in the Monte Carlo generator
with which one simulates the events predicted in the experiment
without and with neutrino oscillations.
However,
since we do not have access to the MiniBooNE Monte Carlo generator,
we adopted an approach which allows the treatment of the multinucleon emission channel,
as well as the quasielastic and the pion production channels through the theoretical model of Ref.~\cite{Martini:2009uj}.
This model has been successful \cite{Martini:2009uj,Martini:2010ex,Martini:2011wp,Martini:2013sha}
to reproduce the MiniBooNE data on the neutrino \cite{AguilarArevalo:2010zc} and antineutrino \cite{AguilarArevalo:2013hm} quasielastic-like cross sections,
as well as the pion production data measured by MiniBooNE \cite{AguilarArevalo:2010bm}
and the T2K data on muon-neutrino \cite{Abe:2013jth} and electron-neutrino \cite{Abe:2014agb}
inclusive cross sections, as shown in Refs.~\cite{Martini:2014dqa,Martini:2016eec}.
This model is based on the nuclear response functions. The quasielastic response is treated in the random phase approximation (RPA), as illustrated for example in Ref.~\cite{Alberico:1981sz}.
The multinucleon contribution is deduced from the microscopic evaluation of Alberico \textit{et al.} \cite{Alberico:1983zg}
of the role of two particle-two hole (2p-2h) excitations in the inclusive $(e,e')$ transverse response.
This calculation includes the correlation term, the two-body meson exchange currents terms, in particular the one associated with Delta excitation,
and the interference between these quantities. The single pion production is assumed to arise exclusively from the pionic decay of the Delta excitation.
In the nucleus the Delta width is modified by medium effects.
They have been introduced and discussed by Oset and Salcedo in Ref.~\cite{Oset:1987re}.
The non pionic decay of the Delta in the medium, which modifies its width leads to 2p-2h or 3p-3h excitations contributing to the multinucleon excitations.
The parameterization of Ref.~\cite{Oset:1987re} for the Delta width in the nuclear medium is used in the model of Ref.~\cite{Martini:2009uj}.

\begin{figure*}
\begin{tabular}{cc}
\includegraphics*[width=0.48\linewidth]{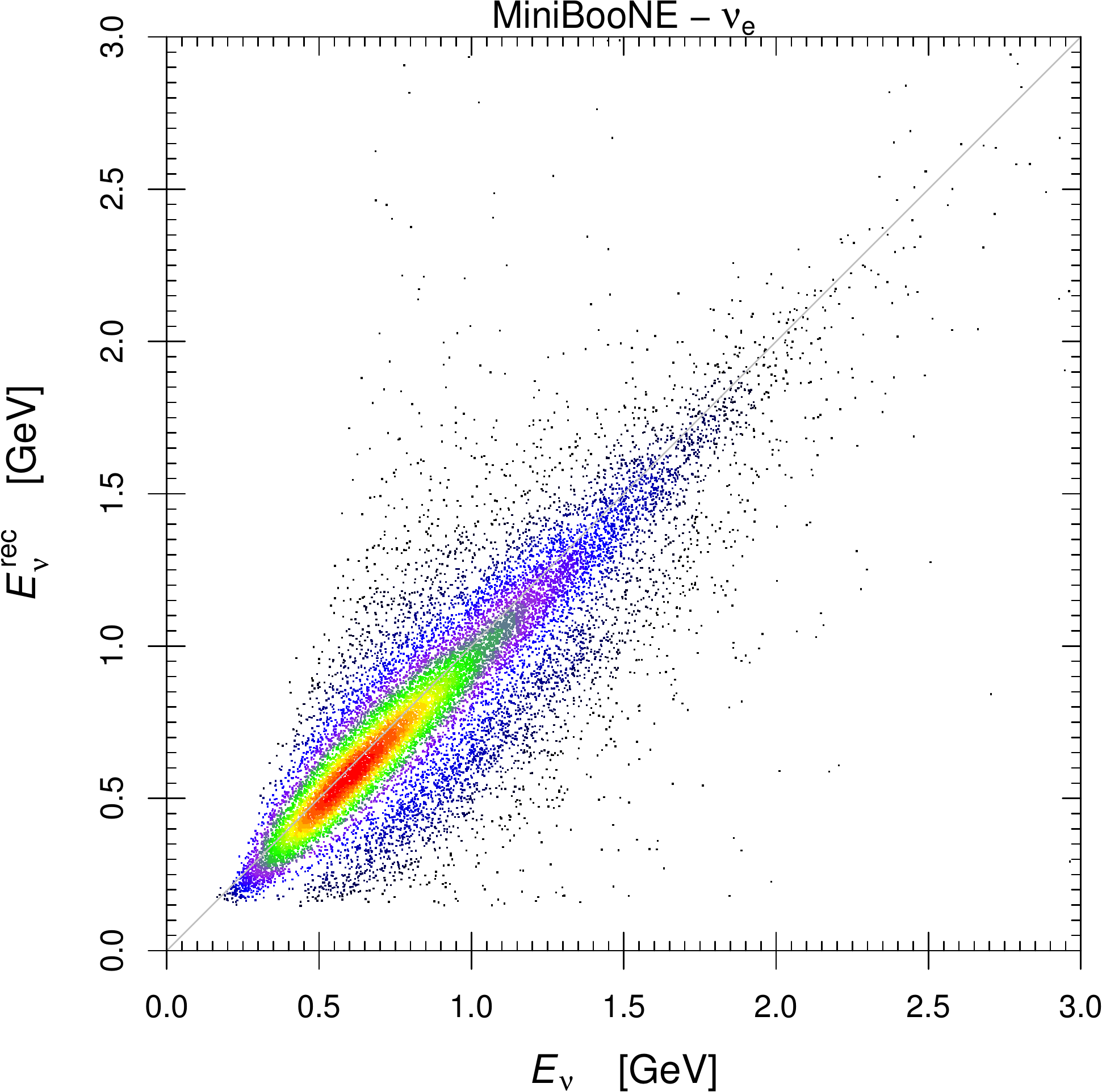}
&
\includegraphics*[width=0.48\linewidth]{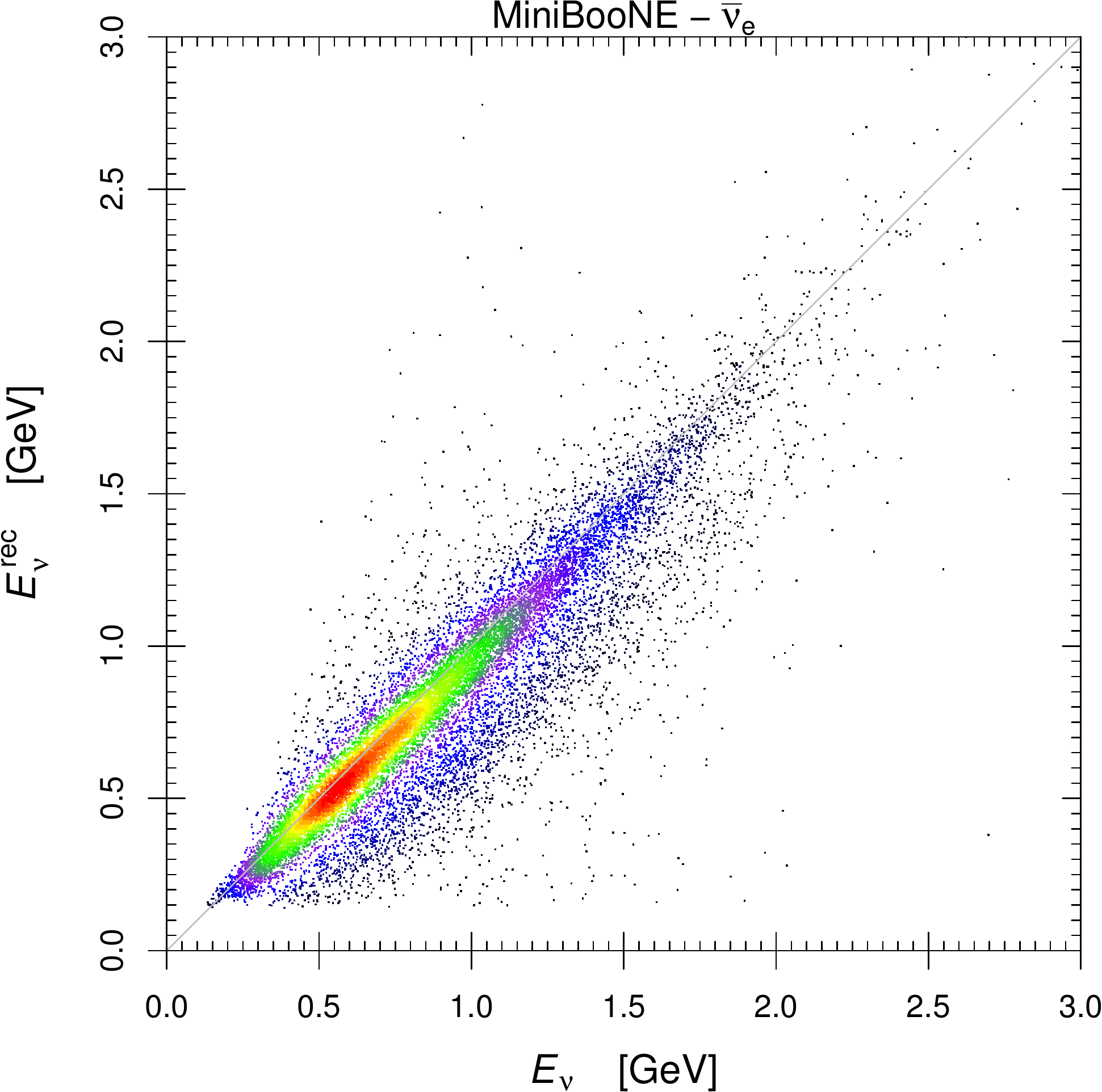}
\\
\includegraphics*[width=0.48\linewidth]{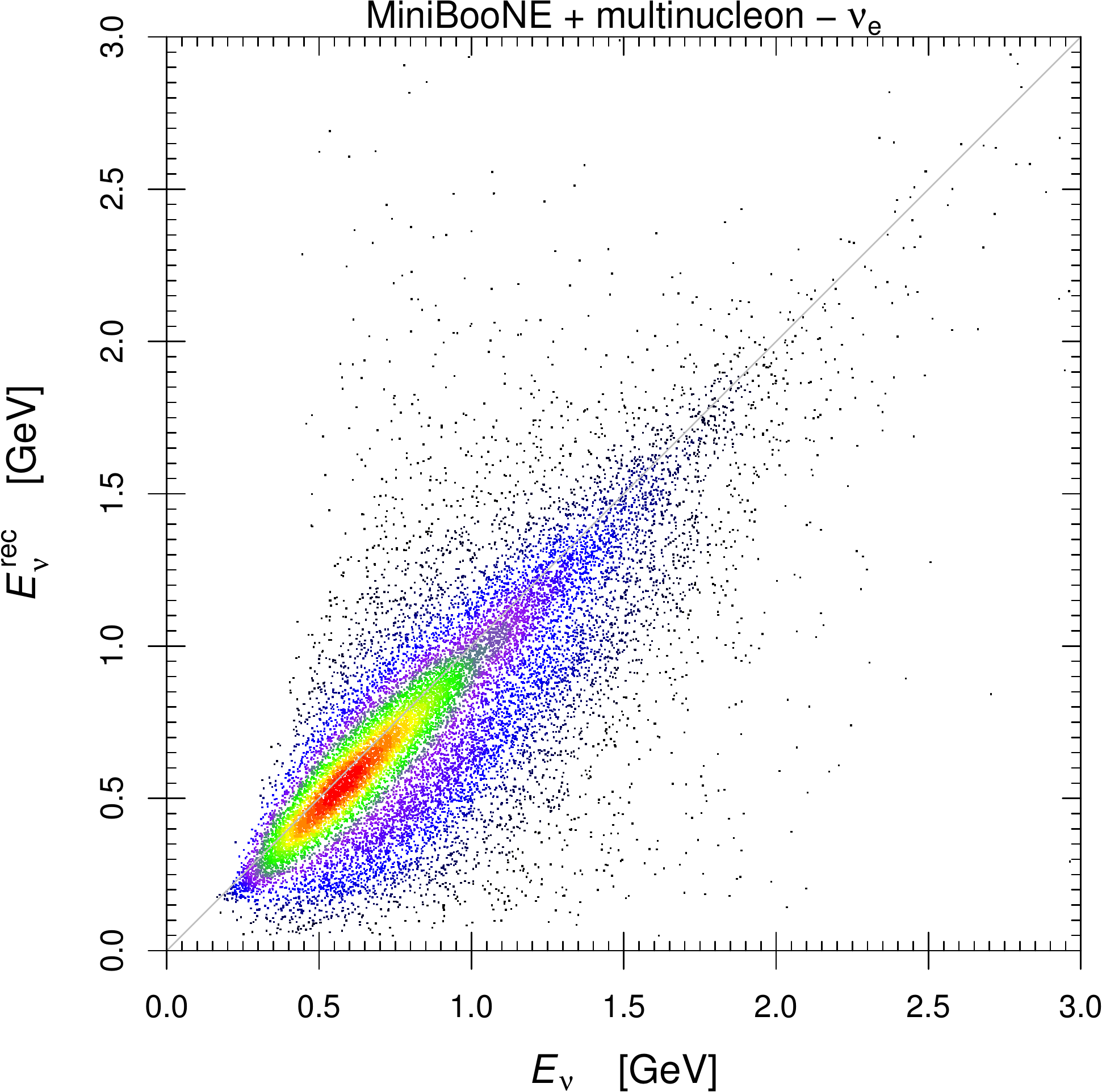}
&
\includegraphics*[width=0.48\linewidth]{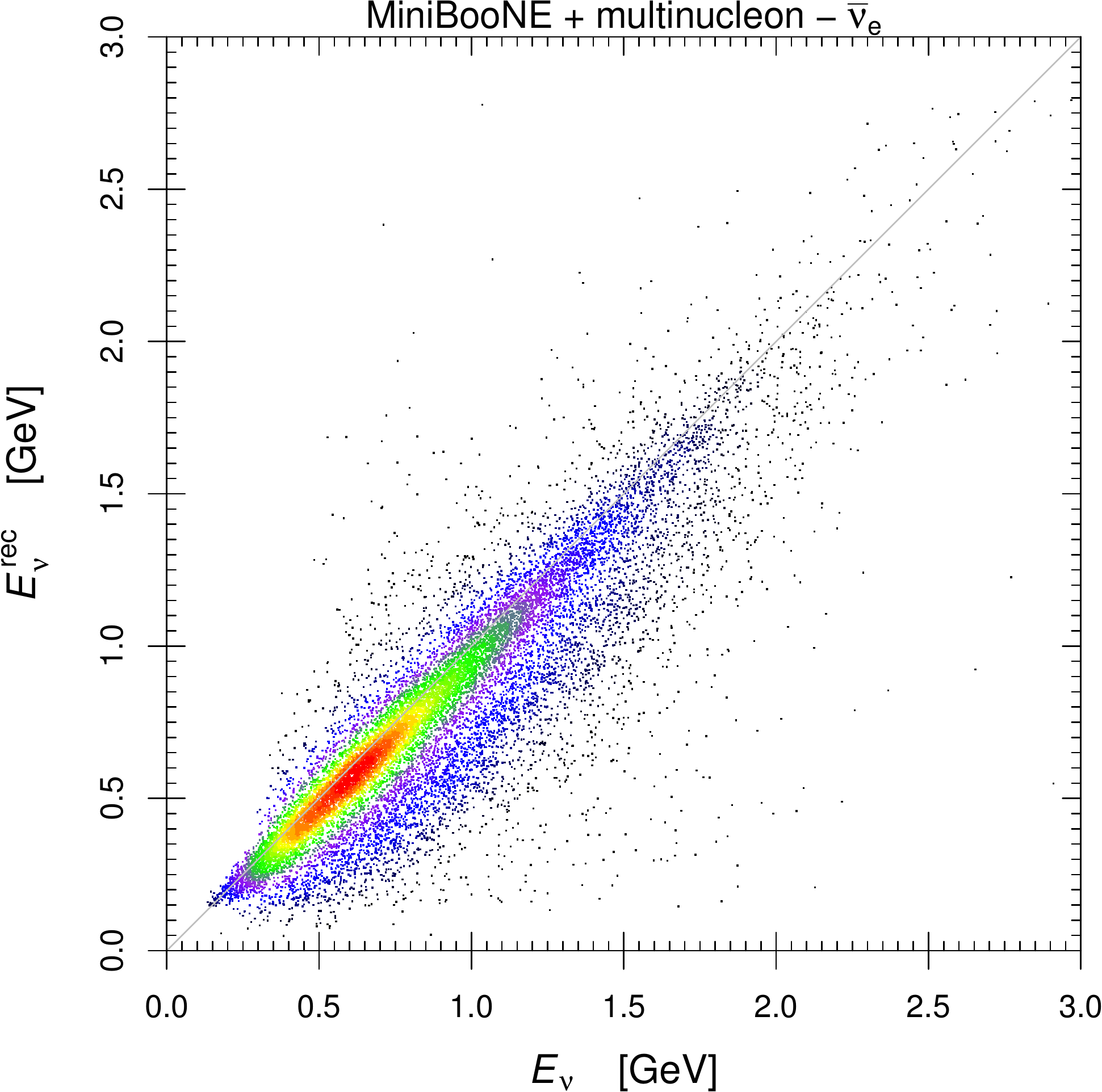}
\end{tabular}
\caption{ \label{fig:scatter}
Scatter plots which shows the correlation between the true neutrino energy $E_{\nu}$
and the reconstructed neutrino energy $E_{\nu}^{\text{rec}}$
in the original
\cite{1207.4809-dr}
MiniBooNE
muon-to-electron neutrino and antineutrino Monte Carlo full transmutation events
(upper plots)
and in the events modified with the contribution of multinucleon interactions
(lower plots).
The color sequence (black, blue, magenta, green, yellow, red)
indicate an increasing density of points.
}
\end{figure*}

\begin{figure*}
\begin{tabular}{ccc}
\includegraphics*[width=0.32\linewidth]{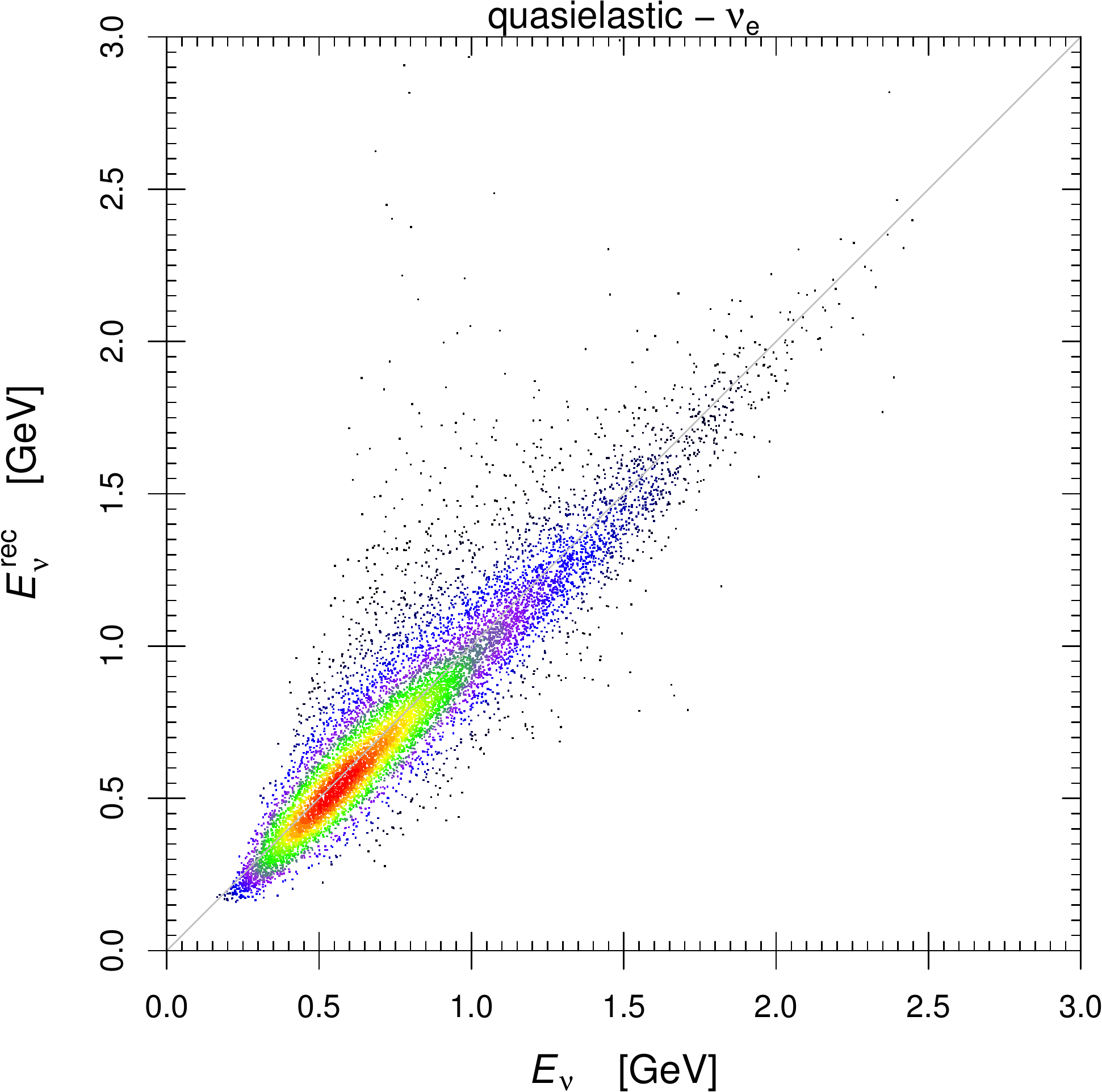}
&
\includegraphics*[width=0.32\linewidth]{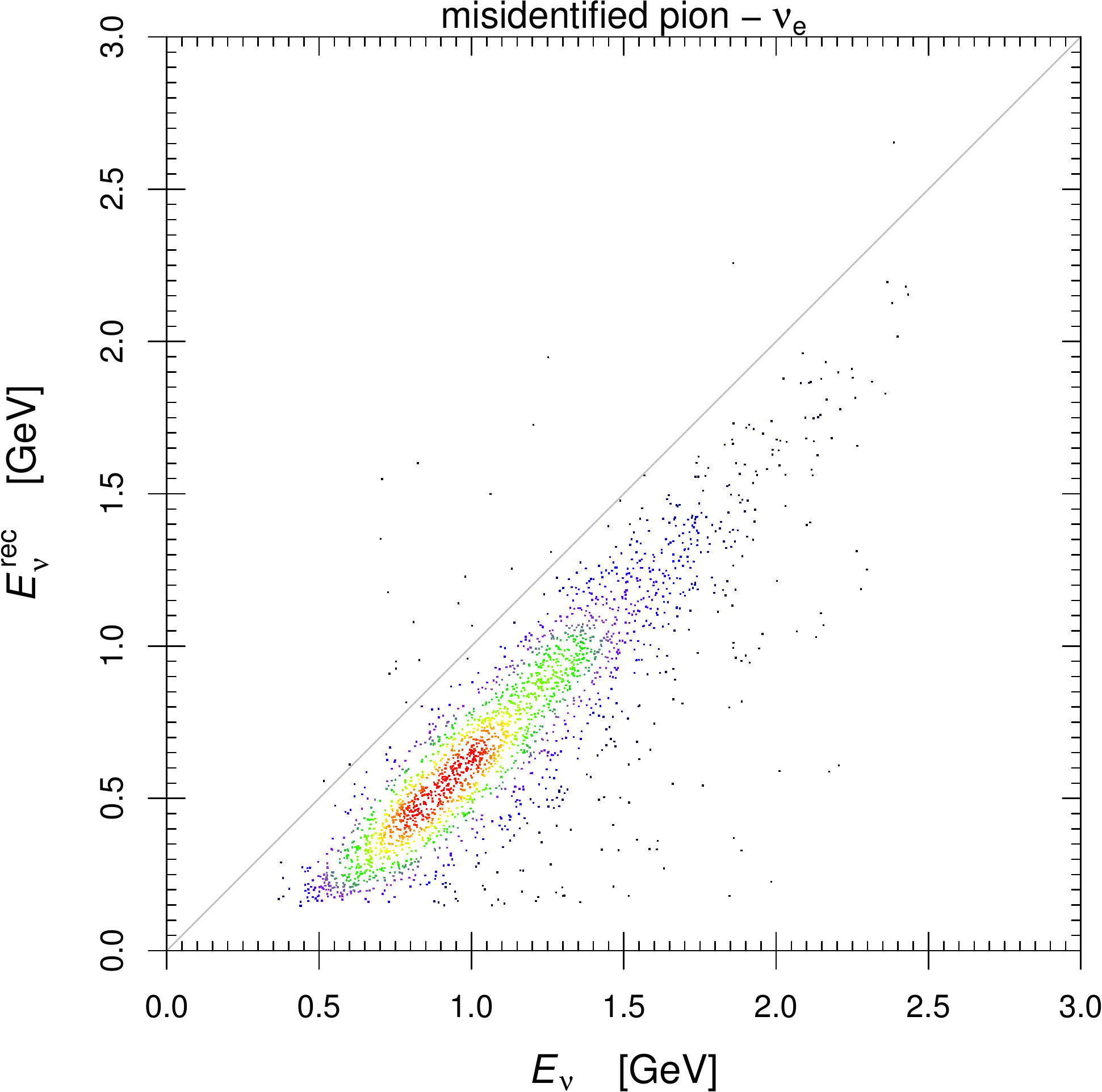}
&
\includegraphics*[width=0.32\linewidth]{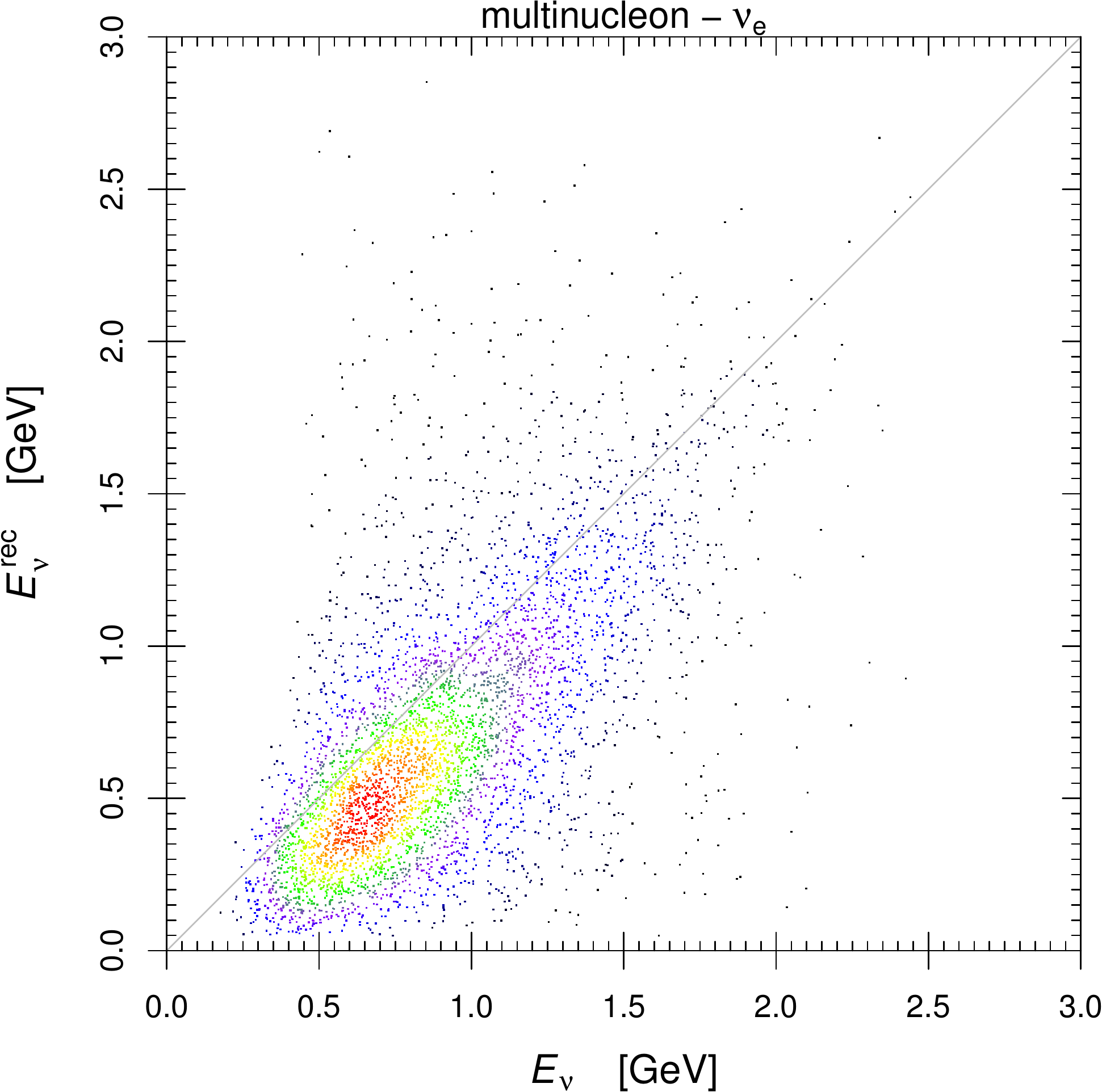}
\\
\includegraphics*[width=0.32\linewidth]{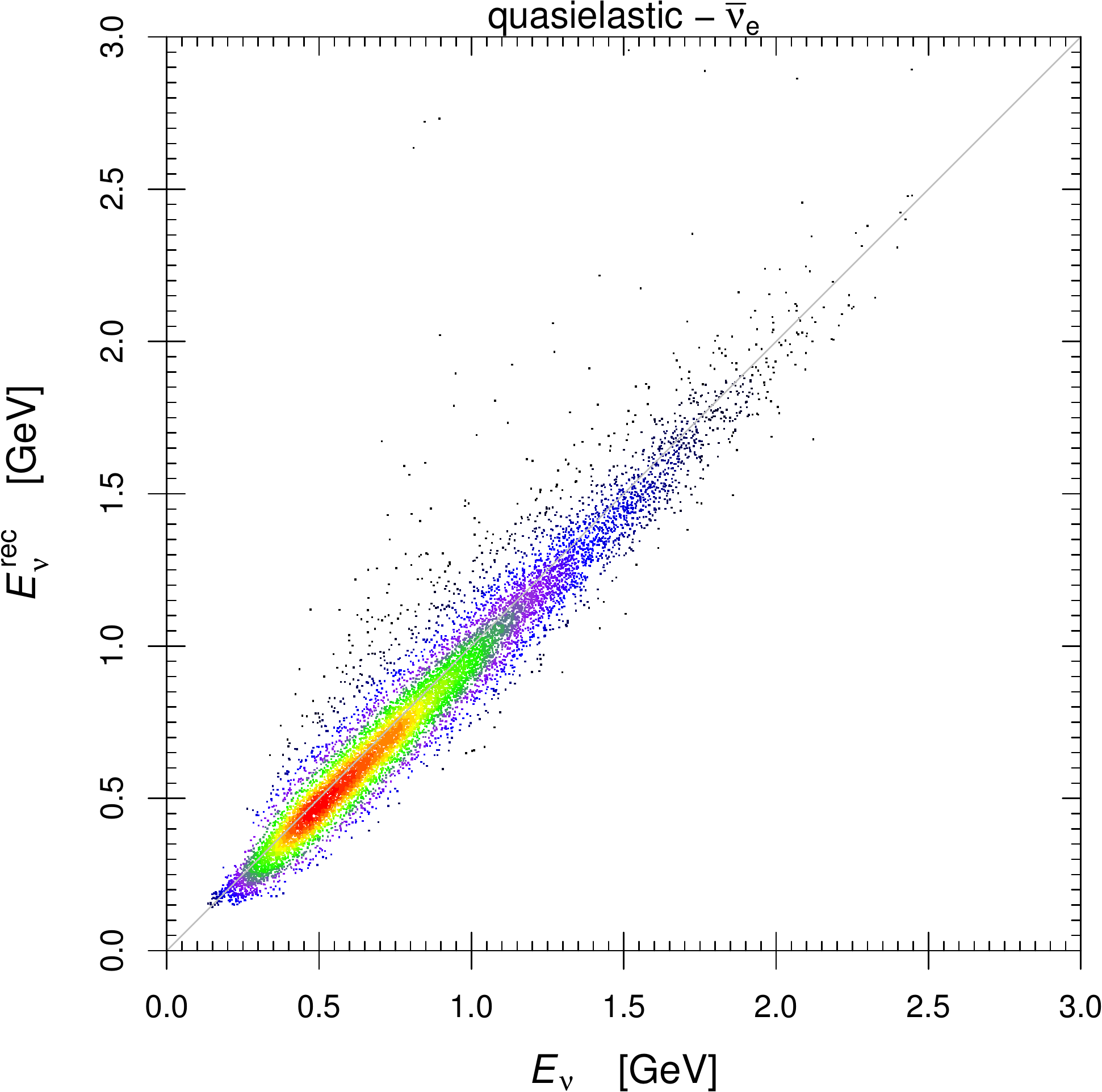}
&
\includegraphics*[width=0.32\linewidth]{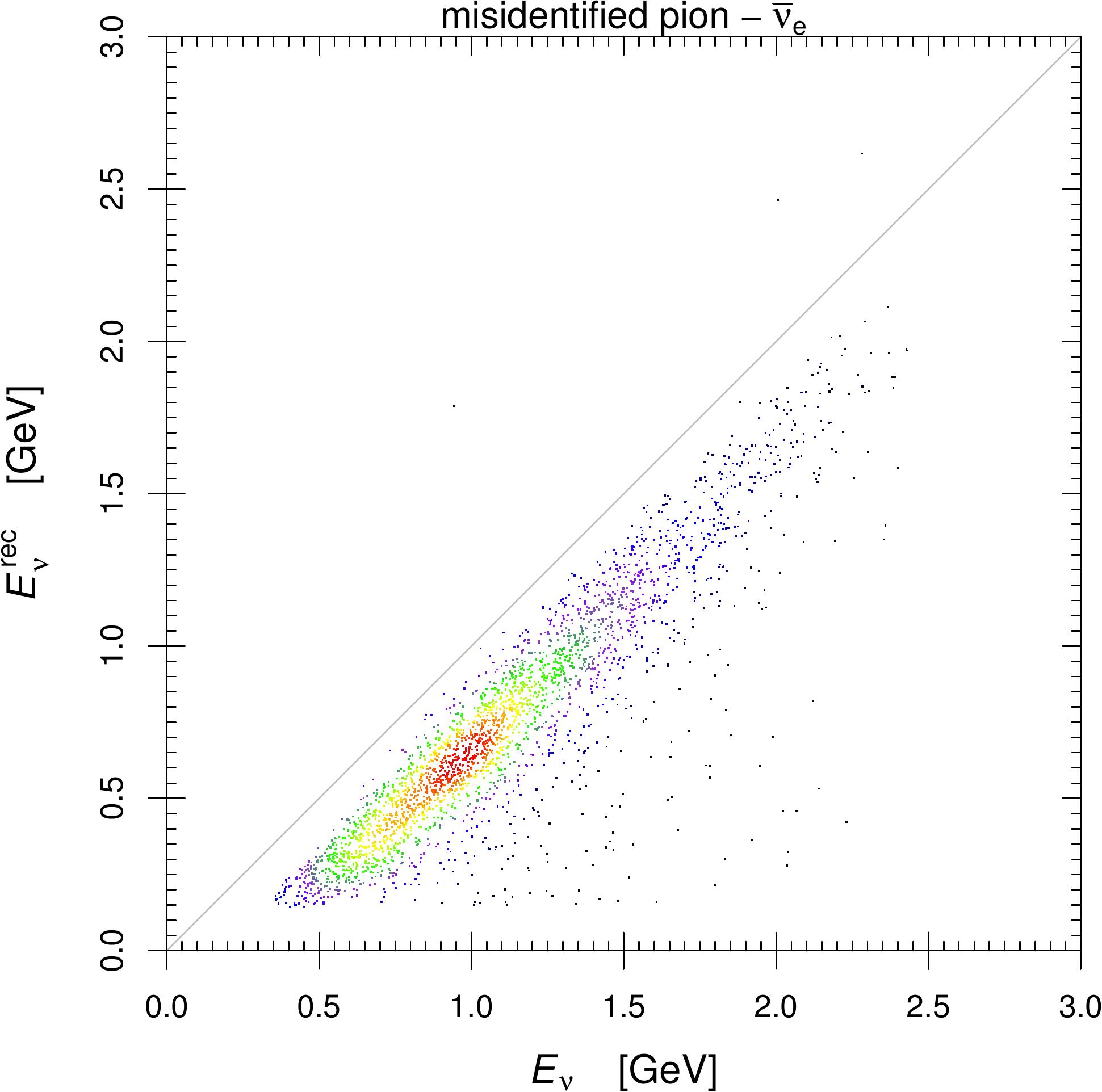}
&
\includegraphics*[width=0.32\linewidth]{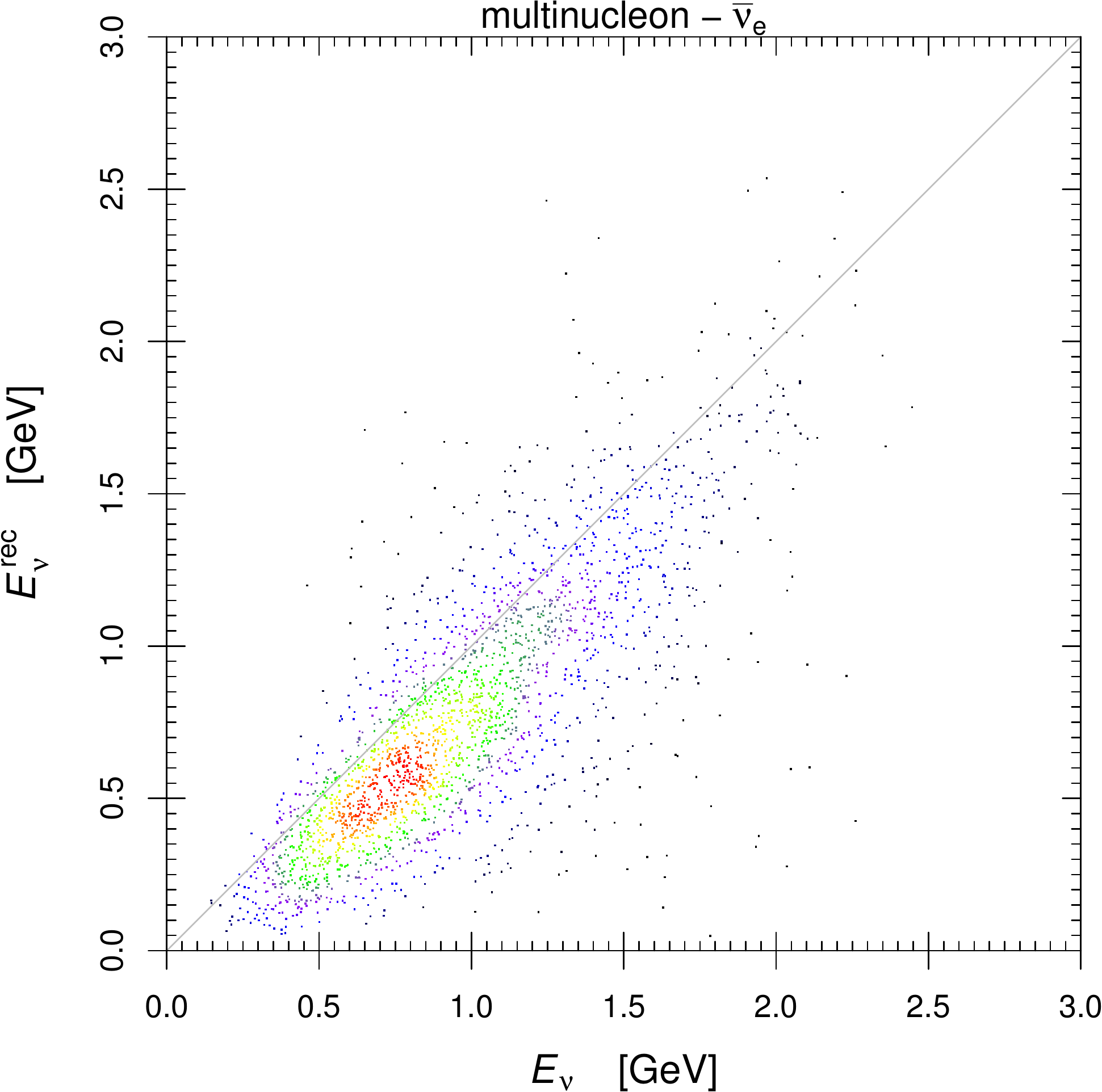}
\end{tabular}
\caption{ \label{fig:qepimn}
Scatter plots which shows the correlation between the true neutrino energy $E_{\nu}$
and the reconstructed neutrino energy $E_{\nu}^{\text{rec}}$
separately
for quasielastic, misidentified pion and multinucleon events
for neutrino
(upper plots)
and antineutrino
(lower plots)
scattering.
The color sequence (black, blue, magenta, green, yellow, red)
indicate an increasing density of points.
}
\end{figure*}

We considered the muon-to-electron neutrino and antineutrino Monte Carlo full transmutation events
in the MiniBooNE data release for the final results of the experiment
\cite{1207.4809-dr}.
The corresponding correlation between the true neutrino energy $E_{\nu}$
and the reconstructed one $E_{\nu}^{\text{rec}}$
is shown in the two upper scatter plots in Fig.~\ref{fig:scatter}.
One can see that most of the points are near the diagonal,
which corresponds to the quasielastic energy reconstruction
\begin{equation}
E_{\nu}^{\text{QE}}
=
\frac{2 \left( M - E_{\text{B}} \right) E_{e} -
      \left( m_{e}^2 - 2 M E_{\text{B}} + E_{\text{B}}^2 + \Delta{M}^2 \right) }
     {2 \left( M - E_{\text{B}} - E_{e} + p_{e} \cos\theta_{e} \right)}
,
\label{QE}
\end{equation}
where $M$ is the mass of the target nucleon which is assumed to be at rest,
$E_{\text{B}} \simeq 25 \, \text{MeV}$ is its binding energy in the nucleus,
$E_{e}$, $p_{e}$ and $\theta_{e}$
are the measured energy, momentum and scattering angle of the outgoing electron
and
$\Delta{M}^2$
is the difference between the squared masses of the initial and final nucleons
($\Delta{M}^2 = M_{n}^2 - M_{p}^2$
for
$\nu_{e} + n \to p + e^{-}$
scattering
and
$\Delta{M}^2 = M_{p}^2 - M_{n}^2$
for
$\bar\nu_{e} + p \to n + e^{+}$
scattering).
The smearing of the quasielastic events around the diagonal
in the two upper scatter plots in Fig.~\ref{fig:scatter}
is due to the Fermi motion of the initial nucleon
and to the electron energy resolution of the detector.
In addition,
one can note an excess of events with reconstructed energy which is significantly
smaller than that in the quasielastic region.
These events are due to charged current pion production
($\nu_{e} + n \to \Delta^{+} + e^{-} \to n + \pi^{+} + e^{-}$ and
$\bar\nu_{e} + p \to \Delta^{0} + e^{+} \to p + \pi^{-} + e^{+}$).
A fraction of the produced pions is not visible
because they are absorbed in the nucleus on their way out.
This process is denoted as the final state interaction effect \cite{Lalakulich:2012cj,Hernandez:2013jka}.
Although its presence has not been displayed \cite{Lalakulich:2012cj,Hernandez:2013jka} in the neutrino-induced charged pion production
MiniBooNE data \cite{AguilarArevalo:2010bm},
this process is present in the cross sections of physical pions \cite{Martini:2009uj} and is expected to be relevant also in the neutrino reactions.
For those pions which do not come out no tracks other than the lepton ones are visible and the process is misidentified as quasielastic event.
In their works on the reconstitution problem \cite{Martini:2012fa,Martini:2012uc},
the authors did not consider the unidentified pions channel but only the quasielastic and the multinucleon channels. In order to exploit their results for the introduction of the multinucleon channel in the MiniBooNE analysis we have
 adopted the following method:

\begin{itemize}

\item
First we separate the quasielastic events from the pion production events. For this,
in the MiniBooNE full transmutation events, we selected statistically the pion production events
which are misidentified as quasielastic charged current events
by choosing the events which have a $E_{\nu}-E_{\nu}^{\text{rec}}$
value which is more likely to be that of
a pion production event than that of a quasielastic event.
The relative probability
of true quasielastic events and misidentified pion production events
is calculated with the nuclear model of Ref.~\cite{Martini:2009uj}.
For the estimation of the relative probability
of misidentified pion production events
we considered\footnote{
We verified that the results do not change in a significant way if we consider
a fraction between 20\% and 40\%.
These percentages are in reasonable agreement
with the indications provided in Ref.~\cite{Lalakulich:2012cj}.
}
30\% of the total charged current single charged pion production events
as misidentified quasielastic charged current events.

We do not apply any change to the selected misidentified pion production events.

\item
We divide randomly the remaining events into
a group which we consider as true quasielastic events
and a group of events which we transform into multinucleon interaction events.

The division is done in proportion to the probability
of quasielastic and multinucleon interactions
calculated in Refs.~\cite{Martini:2012fa,Martini:2012uc}
for different
$(E_{\nu},E_{\nu}^{\text{rec}})$ pairs,
taking into account that
the MiniBooNE detector is filled with pure CH$_2$ mineral oil.
Hence,
in neutrino mode
($\nu_{e} + n \to p + e^{-}$)
all the scatterings can have a multinucleon contribution
because they occur on the neutrons in the carbon atoms,
whereas in antineutrino mode
($\bar\nu_{e} + p \to n + e^{+}$)
only a fraction 3/4 of the scatterings are
with the protons of the carbon atoms.

We do not apply any change to the events in the true quasielastic group.

\item
We consider the group of multinucleon interaction events
for which we calculate a new neutrino reconstructed energy
using the theoretical correlation
between the true neutrino energy $E_{\nu}$
and the reconstructed neutrino energy $E_{\nu}^{\text{rec}}$
calculated in Refs.~\cite{Martini:2012fa,Martini:2012uc}.
We also take into account the energy resolution of the detector
given in Fig.~9.19 of Ref.~\cite{Patterson:2007zz}.

\end{itemize}

\begin{figure*}
\begin{tabular}{cc}
\includegraphics*[width=0.48\linewidth]{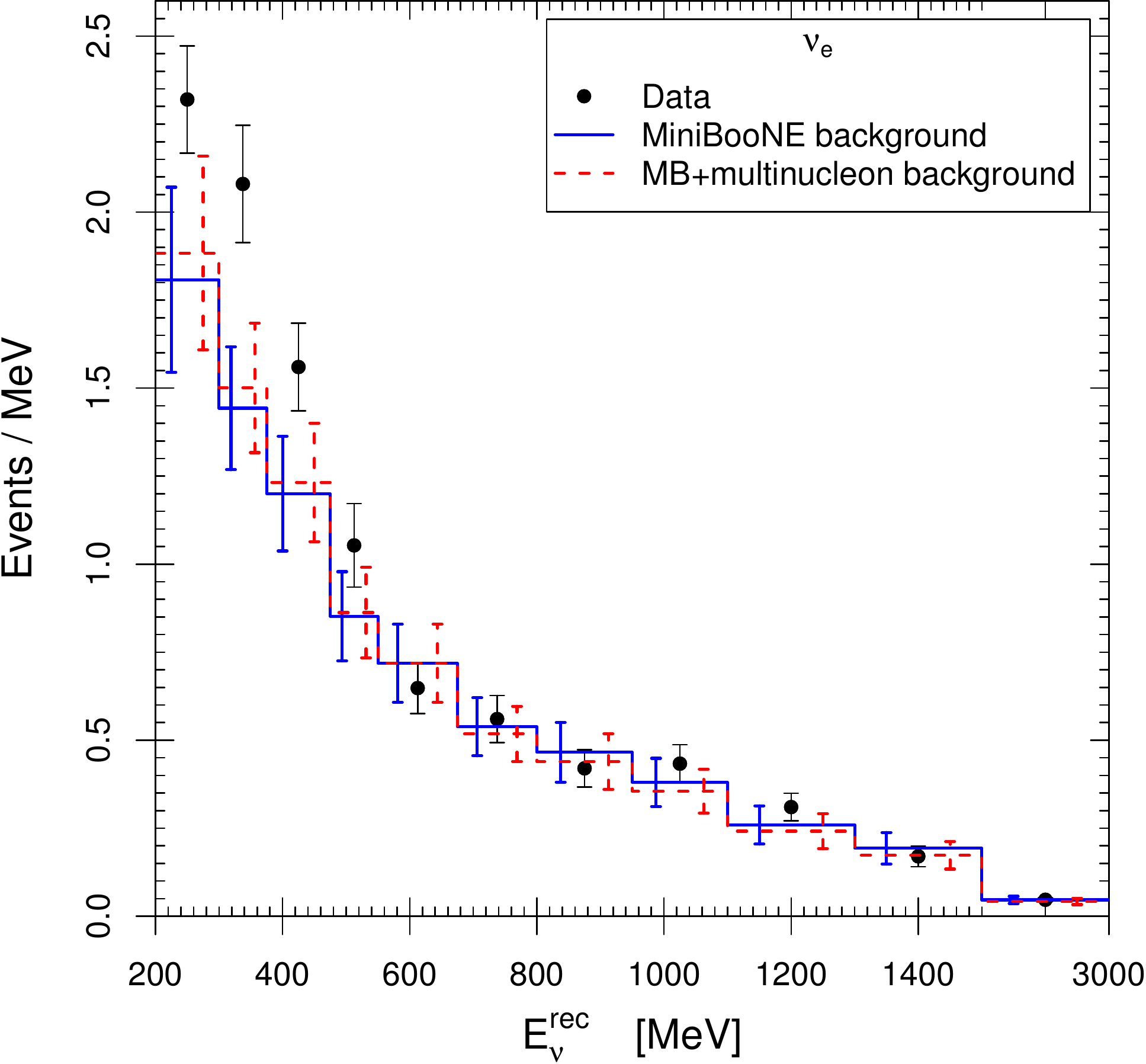}
&
\includegraphics*[width=0.48\linewidth]{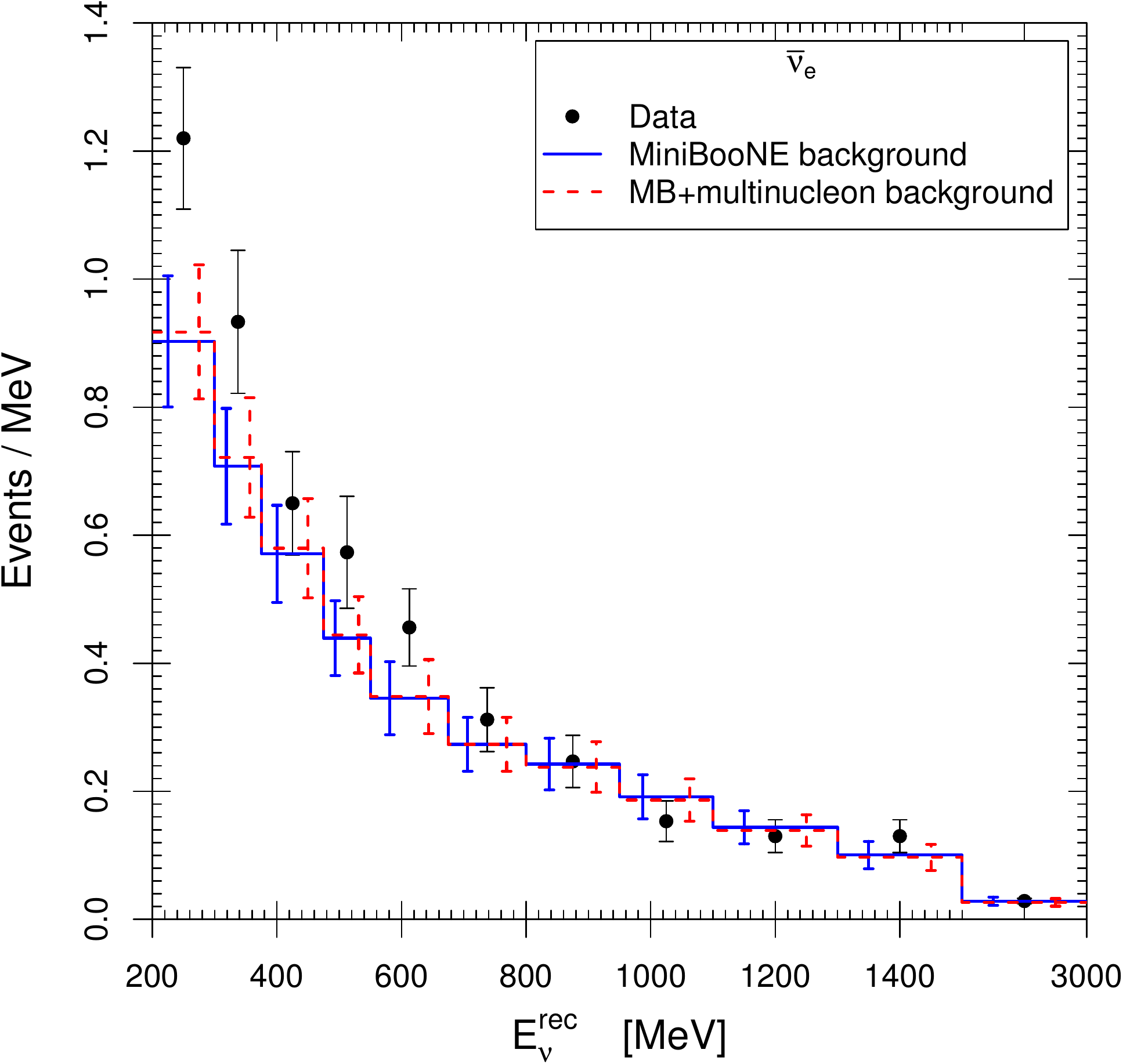}
\end{tabular}
\caption{ \label{fig:eve}
Comparison between the background in the MiniBooNE data release (solid blue)
and that obtained taking into account the multinucleon interactions (dashed red),
with the respective systematic uncertainties.
The left and right figures correspond, respectively, to the neutrino and antineutrino modes.
The data are shown with statistical error bars.
}
\end{figure*}

\begin{table*}
\begin{center}
\renewcommand{\arraystretch}{1.2}
\begin{tabular}{lcc|ccccccccccc|ccccccccccc}
&&&
\multicolumn{11}{c|}{neutrino mode energy bins}
&
\multicolumn{11}{c}{antineutrino mode energy bins}
\\
&
$\sin^22\vartheta$
&
$\Delta{m^2}$
& 1 & 2 & 3 & 4 & 5 & 6 & 7 & 8 & 9 & 10 & 11
& 1 & 2 & 3 & 4 & 5 & 6 & 7 & 8 & 9 & 10 & 11
\\
\hline
MB
&
0
&
0
&
1.7 & 2.6 & 1.8 & 1.2 & -0.5 & 0.2 & -0.5 & 0.6 & 0.8 & -0.4 & 0.04
&
2.1 & 1.6 & 0.7 & 1.3 & 1.3 & 0.6 & 0.07 & -0.8 & -0.4 & 0.9 & 0.03
\\
&
1.0
&
0.04
&
0.8 & 1.5 & 0.3 & -0.3 & -2.0 & -1.1 & -1.3 & 0.03 & 0.3 & -0.7 & -0.07
&
1.0 & 0.5 & -0.5 & 0.3 & 0.4 & -0.2 & -0.5 & -1.2 & -0.7 & 0.7 & -0.03
\\
&
0.01
&
0.4
&
1.0 & 1.6 & 0.5 & -0.1 & -1.9 & -1.0 & -1.3 & 0.05 & 0.3 & -0.7 & -0.07
&
1.3 & 0.7 & -0.3 & 0.4 & 0.5 & -0.2 & -0.5 & -1.2 & -0.7 & 0.7 & -0.03
\\
&
0.003
&
0.7
&
1.3 & 2.0 & 0.8 & 0.1 & -1.7 & -0.8 & -1.2 & 0.1 & 0.4 & -0.6 & -0.06
&
1.7 & 1.0 & -0.04 & 0.6 & 0.6 & -0.05 & -0.4 & -1.2 & -0.6 & 0.8 & -0.03
\\
&
0.003
&
4.0
&
1.4 & 2.3 & 1.1 & 0.002 & -1.9 & -0.7 & -1.2 & -0.5 & -0.8 & -1.8 & -0.9
&
1.9 & 1.2 & 0.2 & 0.4 & 0.4 & 0.1 & -0.4 & -1.7 & -1.7 & -0.2 & -0.6
\\
\hline
MB
&
0
&
0
&
1.4 & 2.3 & 1.6 & 1.1 & -0.5 & 0.4 & -0.2 & 0.9 & 1.1 & -0.06 & 0.5
&
2.0 & 1.5 & 0.6 & 1.2 & 1.3 & 0.6 & 0.2 & -0.7 & -0.3 & 1.0 & 0.3
\\
+
&
0.98
&
0.04
&
0.4 & 1.1 & 0.2 & -0.3 & -1.9 & -0.8 & -1.0 & 0.4 & 0.7 & -0.3 & 0.3
&
0.8 & 0.4 & -0.6 & 0.2 & 0.4 & -0.2 & -0.4 & -1.1 & -0.5 & 0.9 & 0.2
\\
m.
&
0.01
&
0.4
&
0.6 & 1.3 & 0.3 & -0.1 & -1.8 & -0.7 & -1.0 & 0.4 & 0.7 & -0.3 & 0.3
&
1.1 & 0.6 & -0.4 & 0.3 & 0.5 & -0.1 & -0.4 & -1.1 & -0.5 & 0.9 & 0.2
\\
&
0.003
&
0.7
&
1.0 & 1.6 & 0.7 & 0.1 & -1.6 & -0.6 & -0.9 & 0.5 & 0.7 & -0.3 & 0.3
&
1.6 & 0.9 & -0.1 & 0.5 & 0.6 & -0.02 & -0.3 & -1.1 & -0.5 & 0.9 & 0.3
\\
&
0.003
&
4.0
&
1.1 & 1.9 & 0.9 & 0.006 & -1.8 & -0.6 & -1.1 & -0.2 & -0.3 & -1.3 & -0.4
&
1.7 & 1.1 & 0.1 & 0.4 & 0.4 & 0.07 & -0.4 & -1.7 & -1.5 & -0.002 & -0.3
\\
\hline
\end{tabular}
\end{center}
\caption{ \label{tab:pulls}
Table of pulls of
the 11 MiniBooNE energy bins
in neutrino mode
(see Fig.~\ref{fig:eve}-left)
and
those in antineutrino mode
(see Fig.~\ref{fig:eve}-right)
obtained without (MB) and with (MB+m.)
the multinucleon interactions,
without neutrino oscillations
($\sin^22\vartheta=\Delta{m}^2=0$)
and with oscillations
for the best-fit values
and for the selected values of the
oscillation parameters
$\sin^2 2\vartheta$ and $\Delta{m}^2$
considered in Section~\ref{two}
(see Figs.~\ref{fig:hst} and Fig.~\ref{fig:cnt}).
}
\end{table*}

\begin{table}
\begin{center}
\begin{tabular}{ll|c|c}
&
&
MiniBooNE
&
MB + multinucleon
\\
\hline
No Osc.	&$\chi^{2}$		&56.0	&54.7	\\
	&NDF			&38	&38	\\
	&GoF			& 3\%& 4\% \\
\hline
Osc.	&$\chi^{2}_{\text{min}}$&37.3		&36.6		\\
	&NDF			&36		&36		\\
	&GoF			&41\%	&44\%	\\
	&$\sin^22\vartheta$	&1.0		&0.98		\\
	&$\Delta{m}^2$		&0.040		&0.041		\\
\hline
\end{tabular}
\end{center}
\caption{ \label{tab:chi2nu}
Results of fit of MiniBooNE data without (MiniBooNE) and with (MB + multinucleon)
the multinucleon interactions,
without neutrino oscillations
(No Osc.)
and with neutrino oscillations
(Osc.)
in the simplest framework of two-neutrino mixing.
}
\end{table}

\begin{figure*}
\begin{tabular}{cc}
\includegraphics*[width=0.48\linewidth]{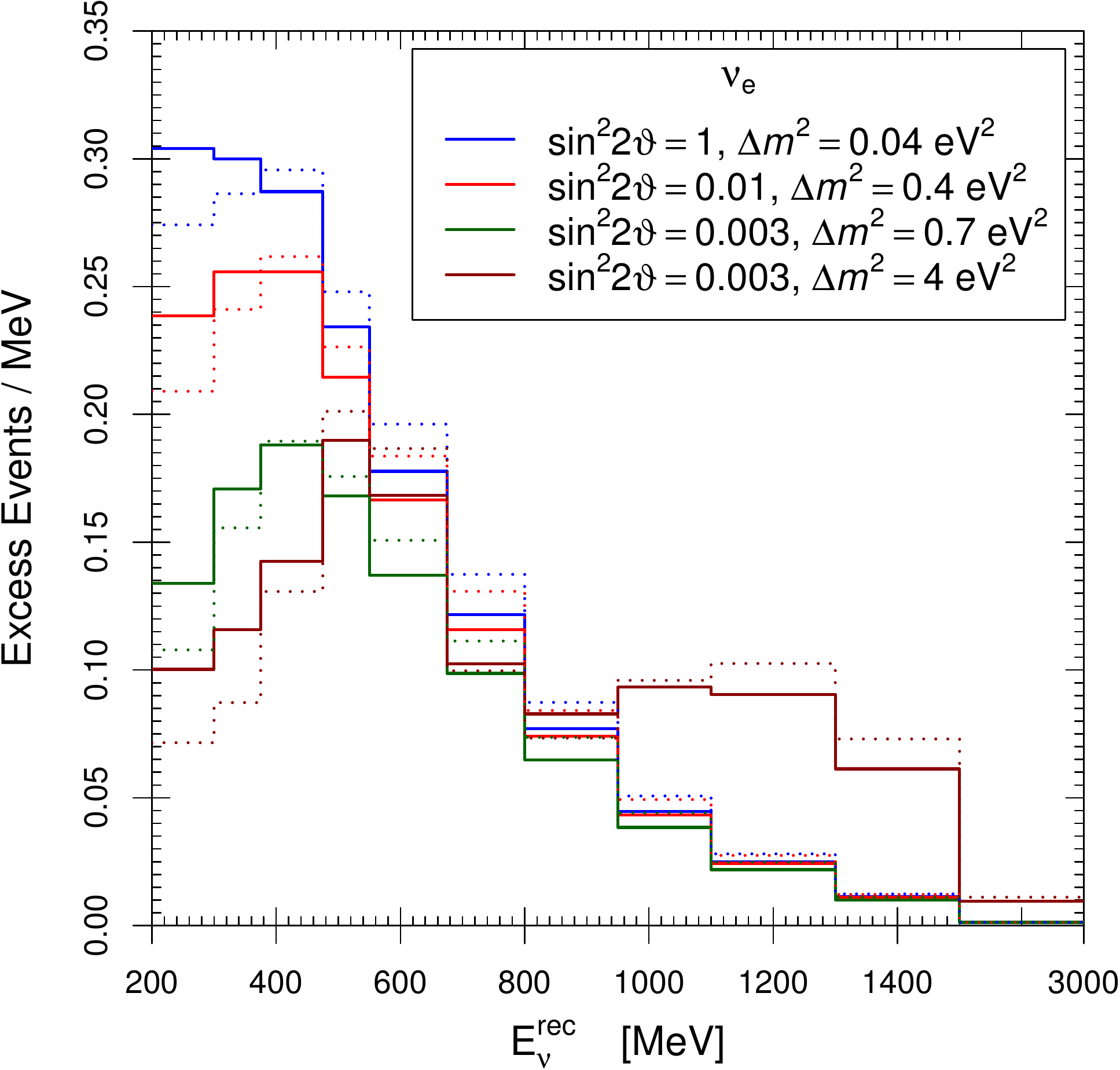}
&
\includegraphics*[width=0.48\linewidth]{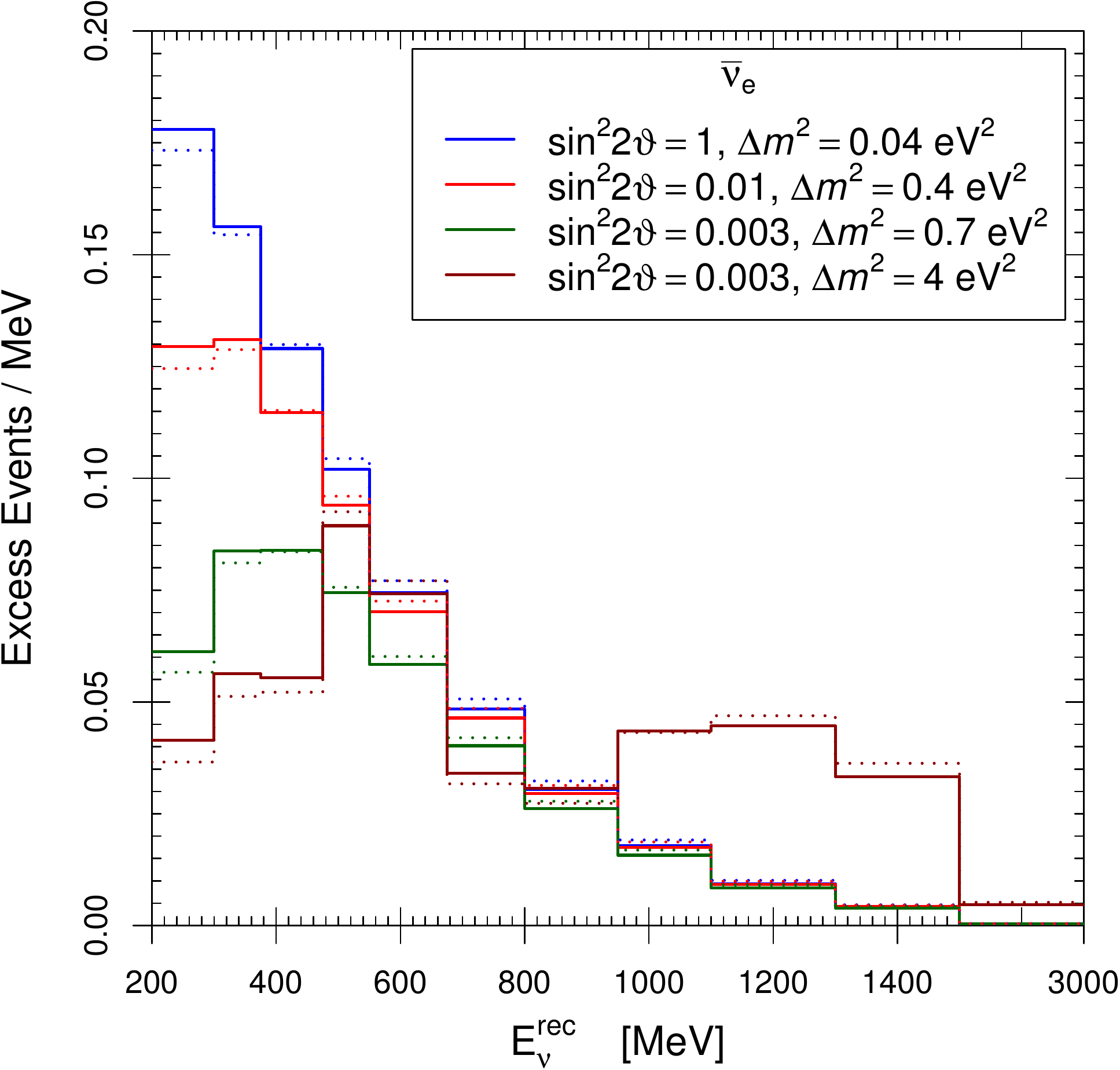}
\end{tabular}
\caption{ \label{fig:comp}
Comparison of the theoretical excess over the background
of $\nu_{e}$ and $\bar\nu_{e}$ events in MiniBooNE
obtained with neutrino oscillations for
some selected values of the
oscillation parameters
$\sin^2 2\vartheta$ and $\Delta{m}^2$
without (dotted lines)
and
with (solid lines)
the multinucleon interactions.
}
\end{figure*}

\begin{figure*}
\begin{tabular}{cc}
\includegraphics*[width=0.48\linewidth]{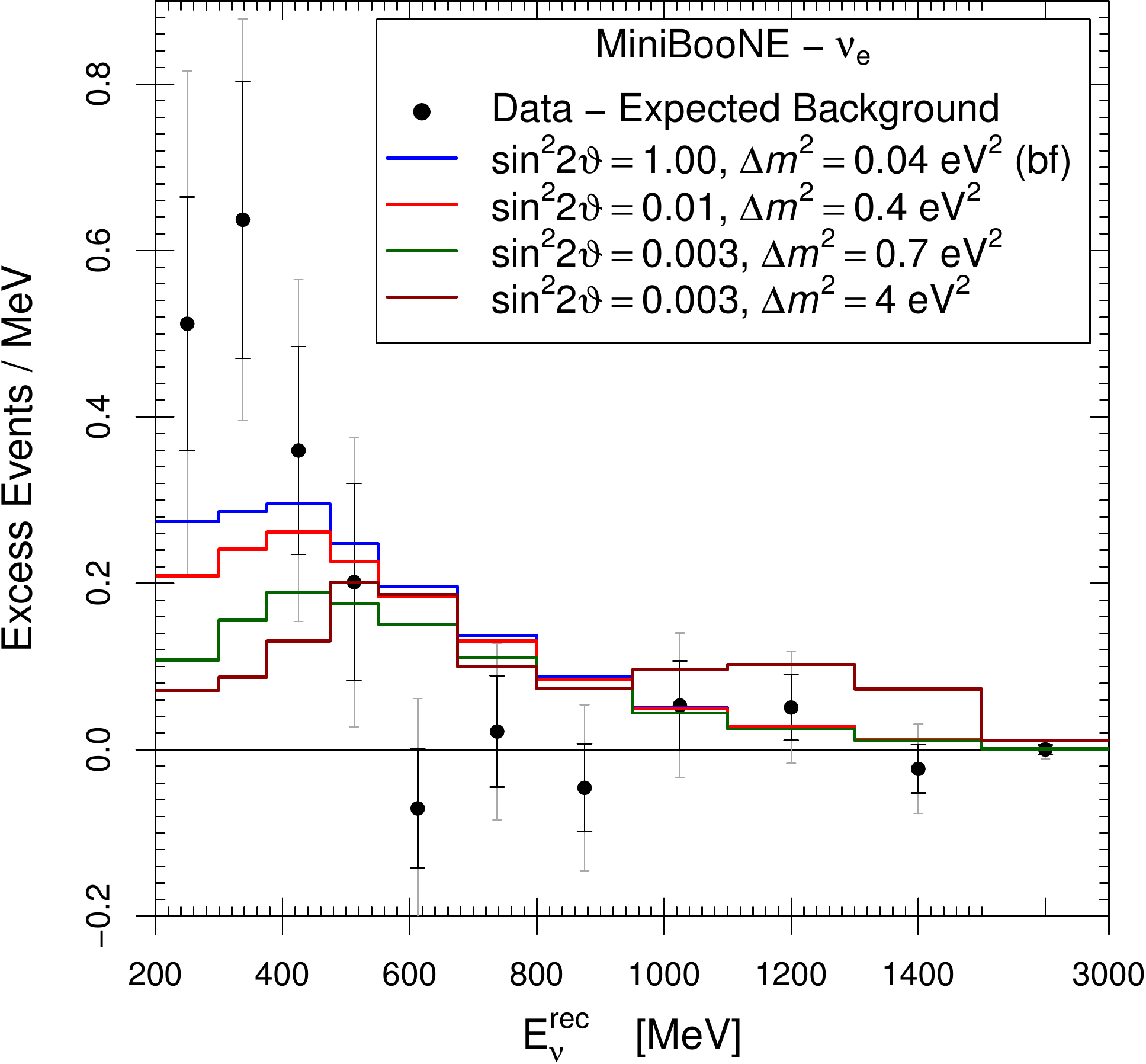}
&
\includegraphics*[width=0.48\linewidth]{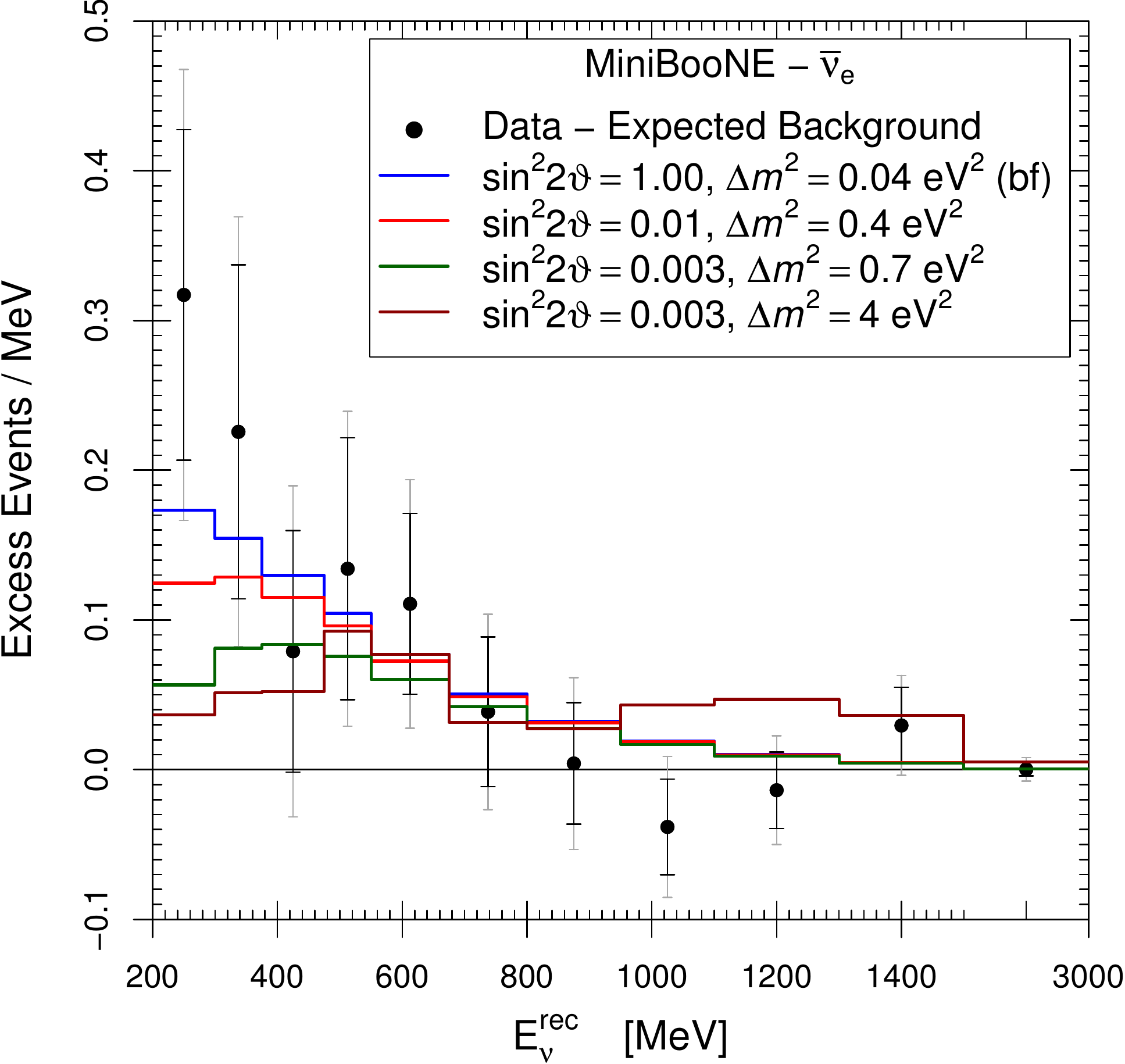}
\\
\includegraphics*[width=0.48\linewidth]{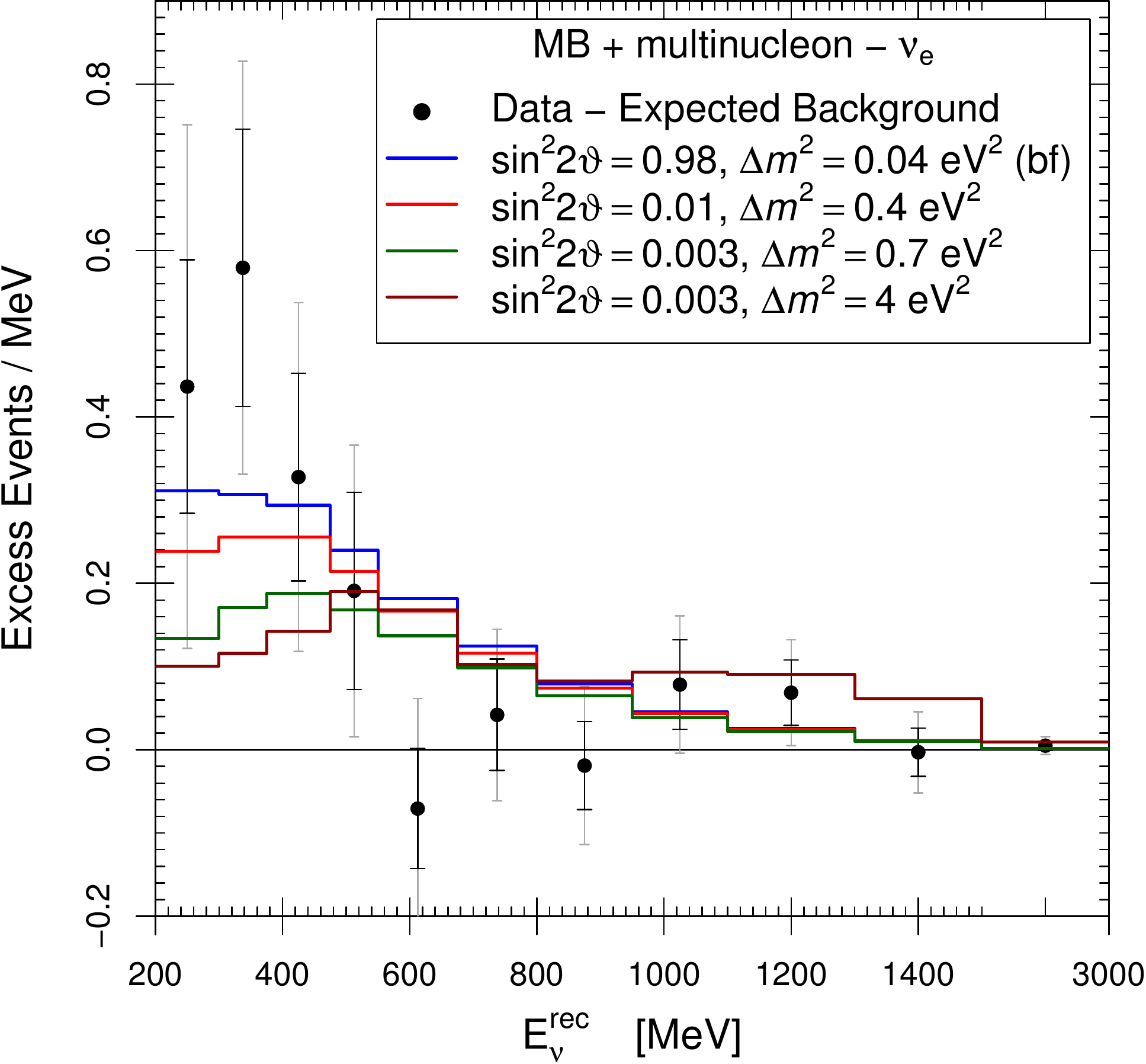}
&
\includegraphics*[width=0.48\linewidth]{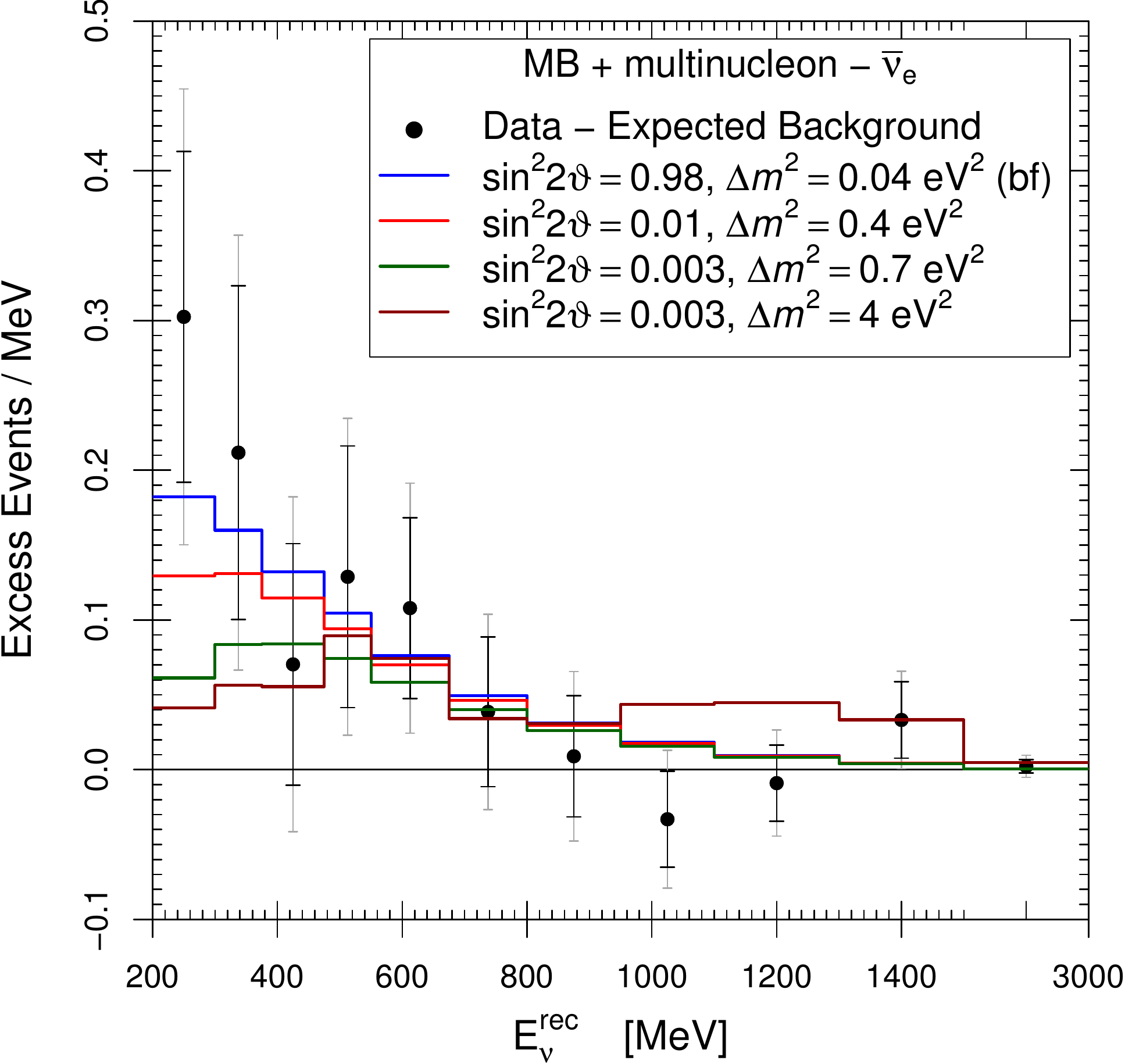}
\end{tabular}
\caption{ \label{fig:hst}
MiniBooNE excess above the background with statistical (black)
and statistical+systematic (gray) error bars in the 11 energy bins
of the MiniBooNE data release in neutrino (left) and antineutrino (right) mode
without (top) and with (bottom) the multinucleon interactions.
The histograms show the predicted excess for
the best-fit values and for some other selected values of the
oscillation parameters
$\sin^2 2\vartheta$ and $\Delta{m}^2$
indicated by the crosses with corresponding color in Fig.~\ref{fig:cnt}.
}
\end{figure*}

\begin{figure*}
\begin{tabular}{cc}
\includegraphics*[width=0.48\linewidth]{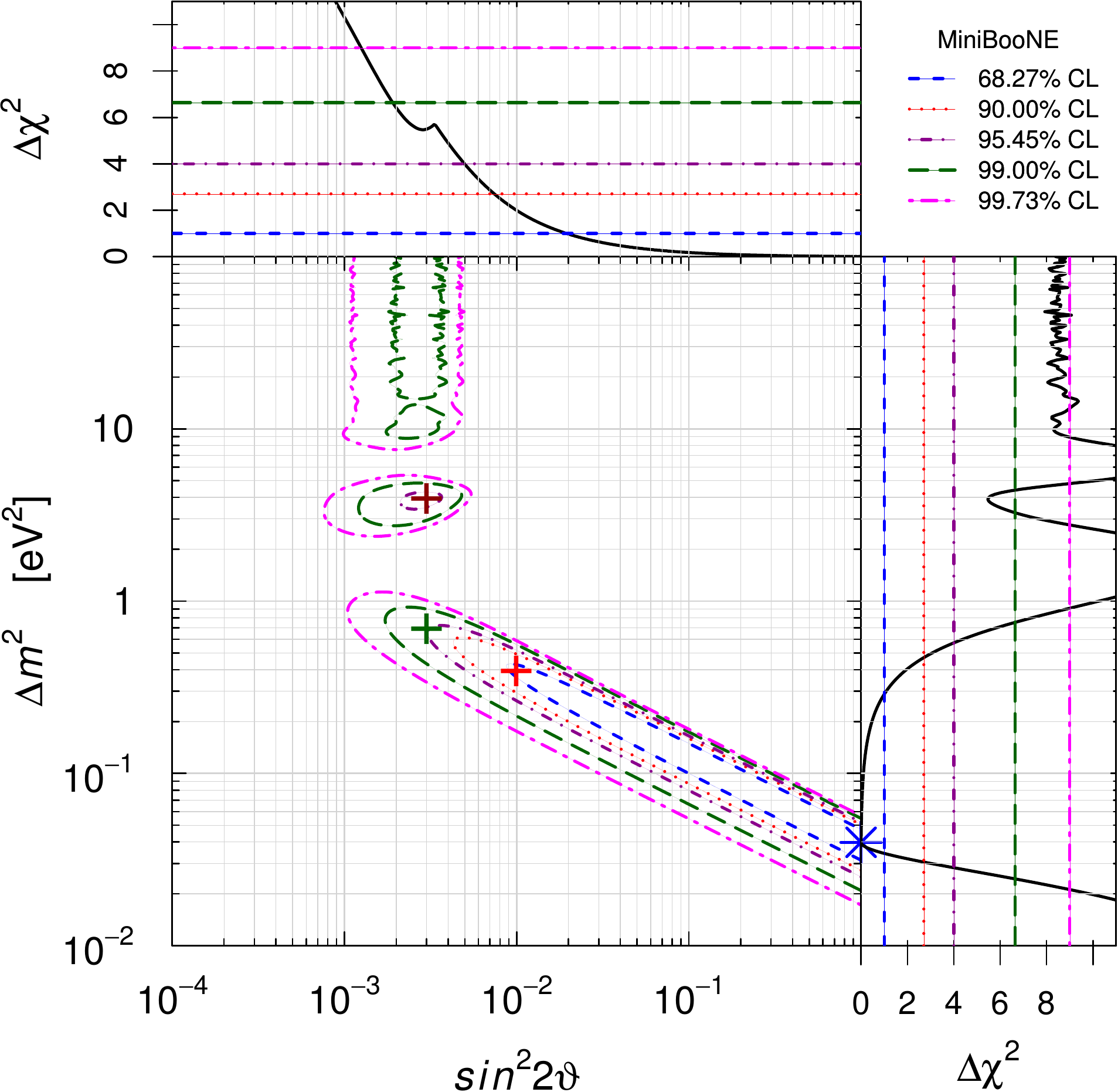}
&
\includegraphics*[width=0.48\linewidth]{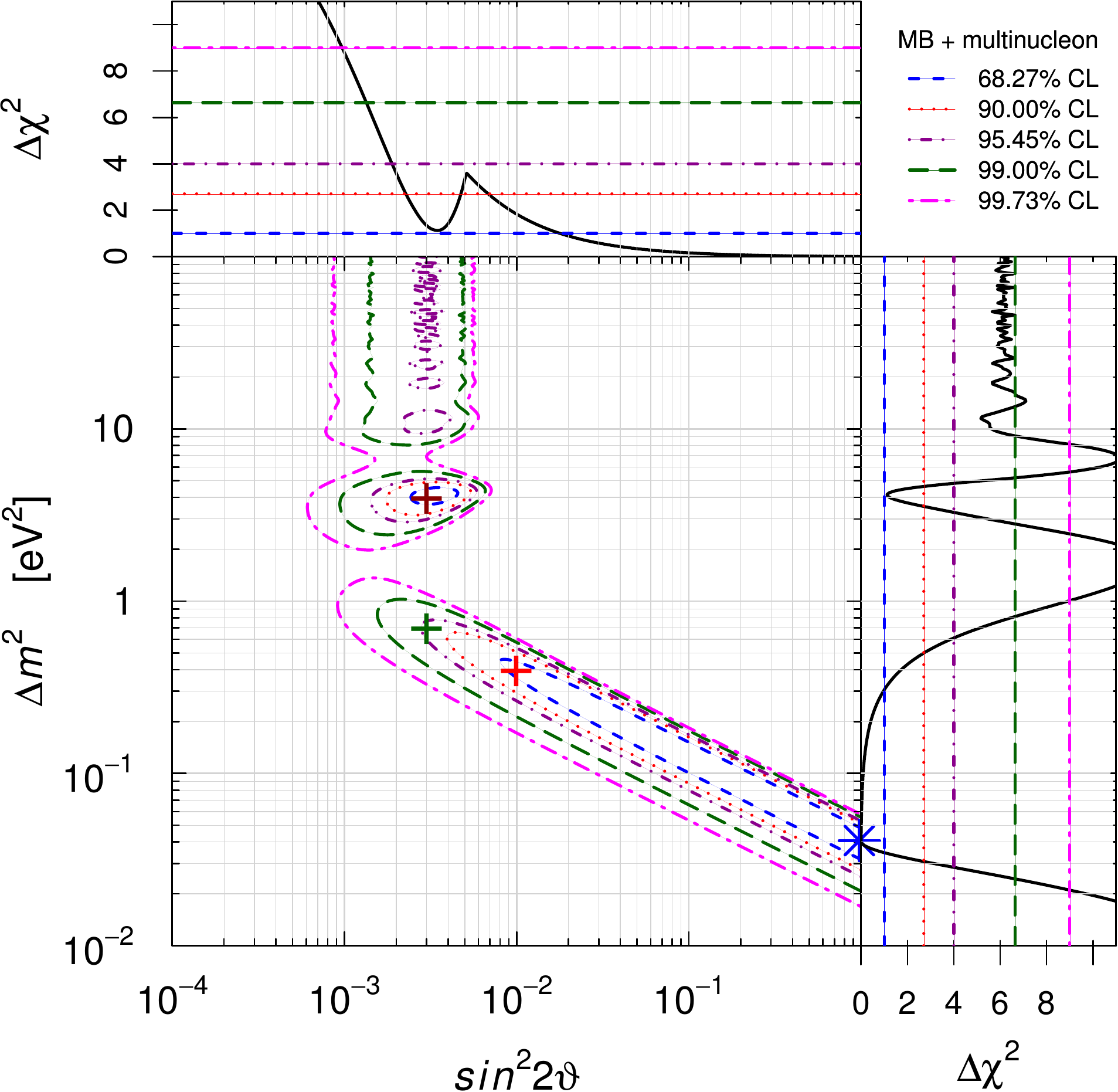}
\end{tabular}
\caption{ \label{fig:cnt}
Allowed regions in the
$\sin^2 2\vartheta$--$\Delta{m}^2$
plane obtained in a two-neutrino mixing fit
of MiniBooNE data without (MiniBooNE) and with (MB + multinucleon)
the multinucleon interactions.
The best-fit point is indicated by a blue asterisk.
The other crosses indicate the values of
$\sin^2 2\vartheta$ and $\Delta{m}^2$
considered in Fig.~\ref{fig:hst}.
}
\end{figure*}

The results of these transformations
are new sets of
neutrino and antineutrino Monte Carlo full transmutation events
with the correlations between the true neutrino energy $E_{\nu}$
and the reconstructed neutrino energy $E_{\nu}^{\text{rec}}$
shown in the two lower scatter plots in Fig.~\ref{fig:scatter}.
Comparing them with the corresponding upper scatter plots
one can see that there are more points
with reconstructed energy which is significantly
smaller than that in the quasielastic region.

Figure~\ref{fig:qepimn} shows the correlation between $E_{\nu}$ and $E_{\nu}^{\text{rec}}$
separately for
quasielastic, misidentified pion and multinucleon events,
for neutrino as well as for antineutrino scattering.
One can notice that the misidentified pion events distribution is displaced horizontally by an amount $\Delta E_{\nu}\simeq300$ MeV
with respect to the quasielastic one, with a smaller $E_{\nu}^{\text{rec}}/E_{\nu}$ ratio.
This shift can be understood by the following argument.
The condition for a quasielastic event on a nucleon at rest is $\omega = E_{\nu} - E_{e} = Q^2/2M_N$,
where $Q^2= |\vet{q}|^2-\omega^2$ is the squared four momentum
and
$M_N$ is the nucleon mass.
For Delta excitation on a nucleon, which is responsible for pion production,
the condition is, neglecting the Delta width,
$\omega = E_{\nu} - E_{e} = Q^2/2M_N+ \Delta M$,
where $\Delta M= (M^2_\Delta- M^2_N)/(2M_N) \simeq 338$ MeV.
This is the amount of the horizontal shift of the misidentified pion production distribution.
As for the distribution of
multinucleon events, it lies between the
distribution of quasielastic events peaked on the diagonal
and
the distribution of misidentified pion events
which have the smallest
$E_{\nu}^{\text{rec}}/E_{\nu}$ ratio.
 The multinucleon interactions
have the effect of shifting some of the events predicted by neutrino oscillations
towards the low energy bins where the anomalous excess is measured.
The effect is larger in neutrino mode,
because in antineutrino mode
the multinucleon interactions are smaller
for two reasons. Firstly the factor 3/4 mentioned
before coming from the nature of the target. Moreover, as was explained in Refs.~\cite{Martini:2010ex,Ericson:2015cva},
in the description of Refs.~\cite{Martini:2009uj,Martini:2010ex} the multinucleon part concerns exclusively the spin-isospin interaction of the weak current
with the nucleus. The corresponding spin-isospin contribution is
reduced for antineutrinos due to the negative value, in this case, of the vector axial interference term.

In the analysis of MiniBooNE data we must also take into account the
evaluation of the background,
which is divided into
$\nua{e}$-induced events
and
misidentified $\nua{\mu}$-induced events.
Since the $\nua{e}$-induced events are produced by the $\nua{e}$'s in the beam generated at the source
by pion and kaon decays,
their interaction is also affected by the multinucleon contribution.
In order to take it into account,
we use the same muon-to-electron neutrino and antineutrino Monte Carlo full transmutation events
considered above
taking into account that they are obtained with the
$\nua{\mu}$ flux and the $\nua{e}$ detection cross section.
In order to transform them into events which correspond to
a $\nua{e}$ flux and a $\nua{e}$ detection cross section
we rescaled the weights of each event according to the ratio of
$\nua{e}$ and $\nua{\mu}$
fluxes in the MiniBooNE beam \cite{AguilarArevalo:2008yp}.
We also normalized the weights of the events in each
reconstructed energy bin in order to reproduce the
number of background $\nua{e}$-induced events
predicted in the MiniBooNE publication
\cite{Aguilar-Arevalo:2013pmq}.
Then, we applied the same method described above in order to add the
effect of multinucleon interactions in the calculated background component
of $\nua{e}$-induced events.
Finally,
we calculate the number of background $\nua{e}$-induced events
taking into account the multinucleon interactions
by summing the weights of the events in each
reconstructed energy bin.

Figure~\ref{fig:eve}
shows the background histogram as a function of $E_{\nu}^{\text{rec}}$
that we have obtained taking into account multinucleon interactions
in comparison with the original
background considered by the MiniBooNE collaboration.
One can see that in neutrino mode the inclusion of the multinucleon interactions
leads to a significant increase of the background
in the low-energy bins
and a small decrease of the background in the high-energy bins.
This effect helps to decrease the low-energy anomaly in neutrino mode, as can be seen from
the pulls in Tab.~\ref{tab:pulls} of the three energy bins below 475 MeV
corresponding to the case of no oscillations
($\sin^22\vartheta=\Delta{m}^2=0$).
In antineutrino mode
the effect is smaller
and there is little change of the pulls
in Tab.~\ref{tab:pulls} of the three energy bins below 475 MeV
corresponding to the case of no oscillations.

The neutrino oscillation analysis of MiniBooNE
is done by using the covariance matrix of systematic uncertainties
given in the MiniBooNE data release
\cite{1207.4809-dr}.
This matrix connects the systematic uncertainties of $\nua{e}$ events in 11 energy bins
to those of $\nua{\mu}$ events in 8 energy bins.
The correlation is important because the $\nua{\mu}$ events,
which are measured with high statistics,
provide a normalization for the $\nua{e}$ event predictions
\cite{AguilarArevalo:2008zz,Karagiorgi:2010zz}.
In principle one should
add the multinucleon interactions also to the analysis
of the $\nua{\mu}$ data.
However,
we could not do it because there is no
publicly available sample of Monte Carlo $\nua{\mu}$ events
corresponding to the final MiniBooNE results in neutrino
\cite{AguilarArevalo:2008rc}
and
antineutrino
\cite{Aguilar-Arevalo:2013pmq}
mode.
Moreover,
the MiniBooNE Monte Carlo has been
tuned in order to fit the
high-statistics $\nua{\mu}$ events
\cite{AguilarArevalo:2008zz,Karagiorgi:2010zz}
through the choices of an overall normalization factor 1.28,
the value of the axial mass
$m_{A} = 1.23 \, \text{GeV}$
and a nuclear Pauli-blocking factor
$\kappa = 1.022$.
Therefore,
our best approach is to neglect possible
corrections to the analysis of $\nua{\mu}$ events
due to multinucleon interactions.

The tuning of the MiniBooNE Monte Carlo through a modification of
the nucleon axial mass parameter in order to fit the measured
$\nu_\mu$ cross sections simulating the multinucleon
influence naturally leads to the following question: is the
multinucleon effect already incorporated in the Monte Carlo $\nu_e$
full transmutation events? In this respect, we can make the
following comment: an increase of the nucleon axial mass
modifies the strength of the response in the quasielastic peak,
but does not extend the region of response beyond this peak,
which is an effect of the multinucleon excitations
\cite{Alberico:1981sz}.
Therefore, it could not
produce the same spreading effect towards smaller reconstructed
neutrino energies as the multinucleon contribution
(right panels in Fig.~\ref{fig:qepimn}).
We can conclude that the multinucleon contribution has to be added.

The first three lines in
Table~\ref{tab:chi2nu}
give the $\chi^2$,
number of degrees of freedom
and
goodness-of-fit
of the fit of MiniBooNE data
without and with multinucleon interactions
in the case of no oscillations.
One can see that the goodness-of-fit is slightly better
when multinucleon interactions are taken into account
but it remains very low.
In the next section we consider
neutrino oscillations
which give a better fit.

\section{Two-neutrino mixing}
\label{two}

First,
we fitted the MiniBooNE data without and with the multinucleon interactions
in the simplest framework of two-neutrino mixing
with the oscillation probability
\begin{equation}
P_{\nua{\mu}\to\nua{e}}
=
\sin^2 2\vartheta \sin^2\left( \frac{\Delta{m}^2 L}{4 E_{\nu}} \right)
,
\label{pemu2nu}
\end{equation}
where $\Delta{m}^2$ is the squared mass difference,
$\vartheta$ is the mixing angle,
$L \simeq 500 \, \text{m}$ is the MiniBooNE source-detector distance
and
$E_{\nu}$ is the neutrino energy.
The results of the fit of MiniBooNE data without and with the multinucleon interactions
are presented in Tab.~\ref{tab:chi2nu}.
One can see that
the multinucleon interactions lead to a decrease of the
$\chi^2$,
as expected from the better fit of the low-energy excess.
However,
the improvement of the fit is rather small
and the best-fit values of
$\sin^2 2\vartheta$
and
$\Delta{m}^2$
obtained from the fits of MiniBooNE data without and with the multinucleon interactions
are similar.
In any case,
there is a significant improvement of the fit
with oscillations with respect to that without oscillations.

Figure~\ref{fig:comp} shows a direct comparison of the theoretical excess over the background
of $\nu_{e}$ and $\bar\nu_{e}$ events in MiniBooNE
obtained with neutrino oscillations for
some selected values of the
oscillation parameters
$\sin^2 2\vartheta$ and $\Delta{m}^2$
without
and
with
the multinucleon interactions.
One can see that in all the cases that we have considered
the inclusion of multinucleon interactions
has the effect of increasing the excess in the low-energy bins,
with small decreases in the high-energy bins.
Hence, we expect that the inclusion of the multinucleon interactions
leads to a better fit of the anomalous low-energy excess.

Figure~\ref{fig:hst}
shows the
excess above the background of the MiniBooNE data in the 11 energy bins
of the MiniBooNE data release in neutrino and antineutrino mode
without and with the multinucleon interactions.
One can see that the low-energy excess slightly decreases
when one takes into account the multinucleon interactions,
with a small increase of the excess in the intermediate and high-energy bins.
This effect is due to the shift shown in Fig.~\ref{fig:eve} of background events
towards lower energies due to the multinucleon interactions.
As already discussed for Fig.~\ref{fig:eve},
the change of the excess due to the multinucleon interactions
is larger in neutrino mode than in antineutrino mode.

The histograms in Fig.~\ref{fig:hst} show the predicted excess for
the best-fit values and for some other selected values of the
oscillation parameters
$\sin^2 2\vartheta$ and $\Delta{m}^2$
which lie inside the allowed regions in the
$\sin^2 2\vartheta$--$\Delta{m}^2$
plane
shown in Fig.~\ref{fig:cnt}.
One can see that
for values of the
oscillation parameters
inside the $1\sigma$ banana-shaped allowed region
that goes from
($\sin^2 2\vartheta \simeq 1$, $\Delta{m}^2 \simeq 0.04$)
to
($\sin^2 2\vartheta \simeq 0.01$, $\Delta{m}^2 \simeq 0.4$)
there is a marginal fit of the low-energy excess in neutrino mode
and this fit is improved when the multinucleon interactions are taken into account
\footnote{
The energy distribution of electron-neutrino events in the target depends on the combined variation of the incident muon flux, the oscillation probability and the electron neutrino cross section in the target. All these quantities are rapidly varying with energy. For instance for $\Delta{m}^2 \simeq 0.4$
the transition probability is maximal for
$E_{\nu} \simeq 200 \, \text{MeV}$,
where
$\Delta{m}^2 L / 2 E_{\nu} \simeq \pi/2$, while the maximum of the energy distribution of the electron events for the same value of $\Delta{m}^2$ occurs at
$E_{\nu} \simeq 400 \, \text{MeV}.$
On the other hand,
for
$\Delta{m}^2 \simeq 0.04$
the transition probability is maximal for
$E_{\nu} \simeq 20 \, \text{MeV}$, while
the peak of the histogram corresponding to
($\sin^2 2\vartheta \simeq 1$, $\Delta{m}^2 \simeq 0.04$)
is also at
$E_{\nu}^{\text{rec}} \simeq 400 \, \text{MeV}$
in neutrino mode without multinucleon interactions.
In this case the excess of electron-neutrino events is also large because in Eq.~(\ref{pemu2nu}) the
smallness of
$\Delta{m}^2 L / 2 E_{\nu}$
is compensated by the large value of
$\sin^2 2\vartheta$.
}.
In antineutrino mode there is little change
and the fit of the low-energy excess is marginally
acceptable in any case.

\begin{figure*}
\begin{tabular}{cc}
\includegraphics*[width=0.48\linewidth]{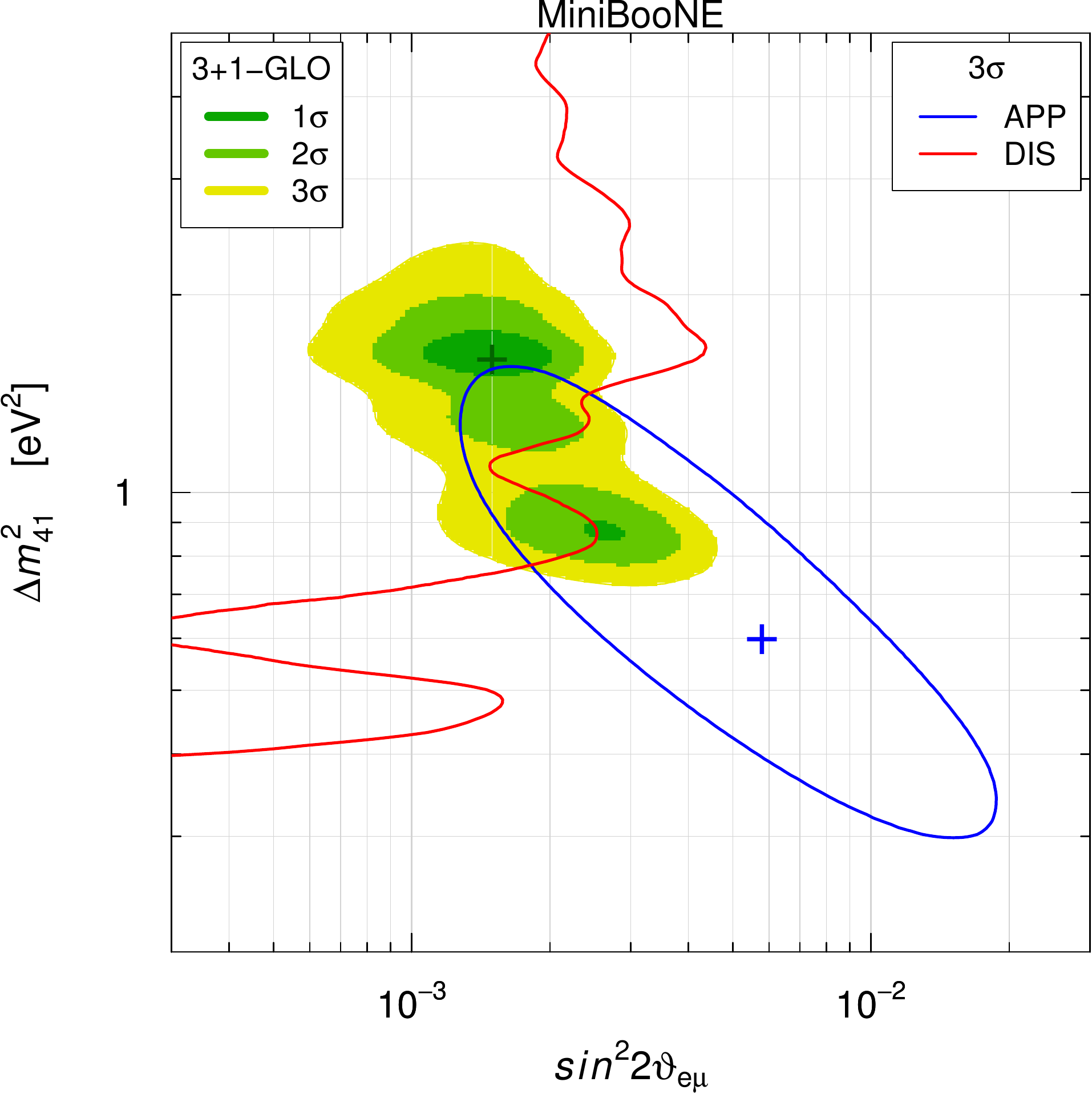}
&
\includegraphics*[width=0.48\linewidth]{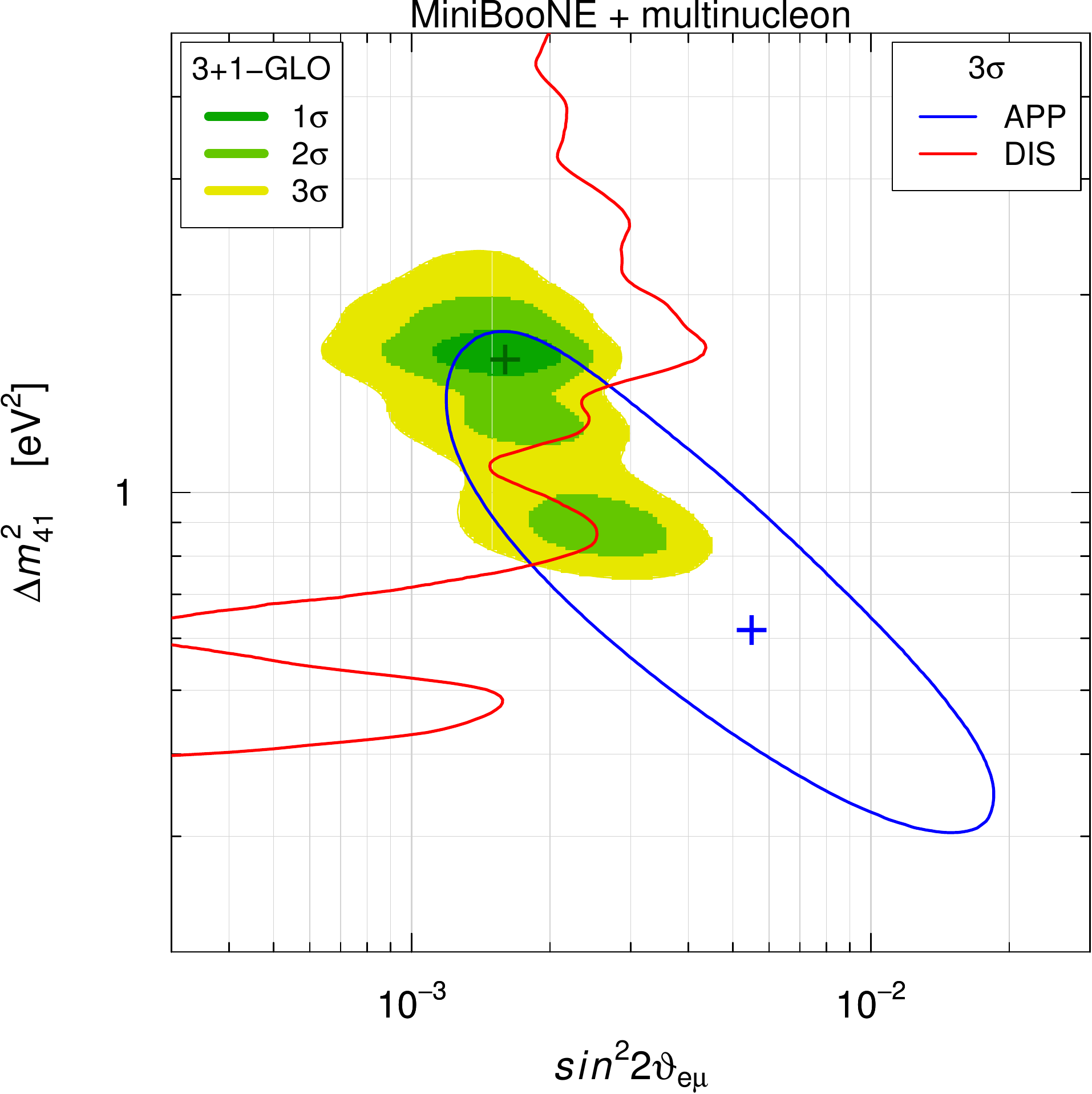}
\end{tabular}
\caption{ \label{fig:3p1}
Allowed regions in the
$\sin^2 2\vartheta_{e\mu}$--$\Delta{m}^2_{41}$
plane obtained in the 3+1 global fit of short-baseline neutrino oscillation data
taking into account the MiniBooNE data without (left) and with (right) the multinucleon interactions.
In both figures
the region inside the blue curve is allowed at $3\sigma$ by
$\protect\nua{\mu}\to\protect\nua{e}$ appearance data, with the best-fit indicated by the blue cross,
the region on the right of the red curve is excluded at $3\sigma$
by $\protect\nua{e}$ and $\protect\nua{\mu}$ disappearance data,
and
the green-yellow shadowed regions
are allowed at $1,2,3\sigma$
by the global fit of all short-baseline neutrino oscillation data.
}
\end{figure*}

However,
values of $\sin^2 2\vartheta$
larger than about 0.003
are excluded by the measurements of several
short-baseline $\nua{e}$ and $\nua{\mu}$ disappearance experiments\footnote{
See the red exclusion curve in Fig.~\ref{fig:3p1},
which has been obtained
in the framework of 3+1 neutrino mixing.
}
which did not observe the large $\nua{e}$ and $\nua{\mu}$
disappearance which must be associated with
the large $\nua{\mu} \to \nua{e}$ transitions given by
$\sin^2 2\vartheta \gtrsim 0.003$
(see Refs.~\cite{Bilenky:1998dt,Strumia:2006db,GonzalezGarcia:2007ib,Abazajian:2012ys,Conrad:2012qt,Palazzo:2013me,Gariazzo:2015rra}).
In fact,
in the simple two-neutrino framework considered so far
we have
$
P_{\nua{e}\to\nua{e}}
=
P_{\nua{\mu}\to\nua{\mu}}
=
1 - P_{\nua{\mu}\to\nua{e}}
$.
A similar constraint holds in any 3+$N_{s}$ neutrino mixing scheme
with three active and $N_{s}$ sterile neutrinos
\cite{Giunti:2015mwa}.

The histograms in Fig.~\ref{fig:hst}
corresponding to
($\sin^2 2\vartheta = 0.003$, $\Delta{m}^2 = 0.7$)
and
($\sin^2 2\vartheta = 0.003$, $\Delta{m}^2 = 4$)
show that when the disappearance constraint is taken into account
the fit of the MiniBooNE low-energy anomaly
is rather bad.
There is a small improvement induced by the consideration of the multinucleon interactions,
but it is not sufficient to produce an acceptable fit,
because the small value of $\sin^2 2\vartheta = 0.003$
gives a probability $P_{\nua{\mu}\to\nua{e}}$ which is too small
and the relatively large values of
$\Delta{m}^2$
lead to a maximum of the event rate which is in the third or a higher energy bin,
whereas the maximum of the low-energy excess is in the first two energy bins.

Let us now discuss the implications of our calculations
for the allowed regions in the
$\sin^2 2\vartheta$--$\Delta{m}^2$
plane shown in Fig.~\ref{fig:cnt},
where we compare the results of the fits of MiniBooNE data
without and with the multinucleon interactions.
One can see that
the multinucleon interactions induce a significant shift
of the allowed regions towards
smaller values of
$\sin^2 2\vartheta$
and
larger values of
$\Delta{m}^2$.
In particular,
the marginal $2\sigma$ lower bound for
$\sin^2 2\vartheta$
changes from
$0.0050$
to
$0.0019$.

This result indicates that the inclusion of the
multinucleon interactions in the analysis of MiniBooNE data
may alleviate the
appearance-disappearance tension
found in the global analyses of the data of
short-baseline neutrino oscillation experiments
\cite{Kopp:2011qd,Giunti:2011gz,Giunti:2011hn,Giunti:2011cp,Conrad:2012qt,Archidiacono:2012ri,Archidiacono:2013xxa,Kopp:2013vaa,Giunti:2013aea,Gariazzo:2015rra,Giunti:2015mwa,Collin:2016rao}.
In fact,
most of this tension
is due to the MiniBooNE low-energy excess,
whose fit requires a small value of $\Delta{m}^2$
and a large value of $\sin^22\vartheta$
\cite{Giunti:2011hn,Giunti:2011cp}.
Decreasing the MiniBooNE low-energy excess
by taking into account the multinucleon interactions
may lead to a significant improvement of the appearance-disappearance tension.

\begin{table}
\begin{center}
\begin{tabular}{l|c|c}
					&MiniBooNE						&MB + multinucleon						\\
\hline
$(\chi^{2}_{\text{min}})_{\text{GLO}}$	&304.0		&300.7			\\
$\text{NDF}_{\text{GLO}}$		&268		&268			\\
$\text{GoF}_{\text{GLO}}$		& 6\%		& 8\%		\\
$\Delta{m}^2_{41}[\text{eV}^2]$		&1.6		&1.6			\\
$|U_{e4}|^2$				&0.028		&0.028			\\
$|U_{\mu4}|^2$				&0.014		&0.015			\\
\hline
$(\chi^{2}_{\text{min}})_{\text{APP}}$	&95.4		&94.0			\\
$(\chi^{2}_{\text{min}})_{\text{DIS}}$	&194.4		&194.4			\\
$\Delta\chi^{2}_{\text{PG}}$		&15.0		&12.7		\\
$\text{NDF}_{\text{PG}}$		&2		&2		\\
$\text{GoF}_{\text{PG}}$		&0.06\%	&0.2\%		\\
\hline
\end{tabular}
\end{center}
\caption{ \label{tab:chi3p1}
Results of the 3+1 global fit of short-baseline neutrino oscillation data
taking into account the MiniBooNE data without (MiniBooNE) and with (MB + multinucleon)
the multinucleon interactions.
The first three lines give
the minimum $\chi^{2}$ ($(\chi^{2}_{\text{min}})_{\text{GLO}}$),
the number of degrees of freedom ($\text{NDF}_{\text{GLO}}$) and
the goodness-of-fit ($\text{GoF}_{\text{GLO}}$)
of the global fit (GLO).
The following three lines
give the best-fit values of
$\Delta{m}^2_{41}$,
$|U_{e4}|^2$ and
$|U_{\mu4}|^2$.
The last five lines give the quantities
relevant for the appearance-disappearance (APP-DIS) parameter goodness-of-fit (PG)
\protect\cite{Maltoni:2003cu}.
}
\end{table}

Let us finally comment on the difference between our results and those presented
by the MiniBooNE collaboration in Ref.~\cite{Aguilar-Arevalo:2013pmq},
which we have mentioned in the Introduction.
Our method for taking into account multinucleon interactions
is an approximate implementation of the correction that one should do
to the Monte Carlo event generator,
which would result in a correction
of the reconstructed neutrino energy $E_{\nu}^{\text{rec}}$ of some events,
without any change of the true neutrino energy $E_{\nu}$ of the events,
which is determined by the neutrino flux.
On the other hand,
according to Ref.~\cite{Garvey:2014exa}
the MiniBooNE collaboration simulated the effect of the multinucleon interactions
``by reassigning the true neutrino energy of some fraction of
events at a given reconstructed neutrino energy to a higher neutrino energy.''
With their procedure,
they obtained a worse fit of the low-energy excess,
as shown in Fig.~36 of Ref.~\cite{Garvey:2014exa},
and a worse best-fit $\chi^2$,
given in Tab.~II of Ref.~\cite{Aguilar-Arevalo:2013pmq},
in contradiction with our results.

\section{Global fit of short-baseline neutrino oscillation data}
\label{3+1}

In order to explore if
the inclusion of the
multinucleon interactions in the analysis of MiniBooNE data
decreases the
appearance-disappearance tension in the global analyses of the data of
short-baseline neutrino oscillation experiments
\cite{Kopp:2011qd,Giunti:2011gz,Giunti:2011hn,Giunti:2011cp,Conrad:2012qt,Archidiacono:2012ri,Archidiacono:2013xxa,Kopp:2013vaa,Giunti:2013aea,Gariazzo:2015rra,Giunti:2015mwa,Collin:2016rao},
we consider the
simplest 3+1 extension of standard three-neutrino mixing
in which there is an additional massive neutrino
$\nu_{4}$
with mass $m_{4} \sim 1 \, \text{eV}$.
In this framework
the effective probability of
$\nua{\alpha}\to\nua{\beta}$
transitions in short-baseline experiments has the two-neutrino-like form
\cite{Bilenky:1996rw}
\begin{equation}
P_{\nua{\alpha}\to\nua{\beta}}
=
\delta_{\alpha\beta}
-
4 |U_{\alpha4}|^2 \left( \delta_{\alpha\beta} - |U_{\beta4}|^2 \right)
\sin^2\!\left( \dfrac{\Delta{m}^2_{41}L}{4E_{\nu}} \right)
,
\label{pab}
\end{equation}
where $U$ is the mixing matrix
and
$\Delta{m}^2_{41} = m_{4}^2 - m_{1}^2 = \Delta{m}^2_{\text{SBL}} \sim 1 \, \text{eV}^2$.
The electron and muon neutrino and antineutrino appearance and disappearance
in short-baseline experiments
depend on
$|U_{e4}|^2$ and $|U_{\mu4}|^2$,
which
determine the amplitude
$\sin^22\vartheta_{e\mu} = 4 |U_{e4}|^2 |U_{\mu4}|^2$
of
$\nua{\mu}\to\nua{e}$
transitions,
the amplitude
$\sin^22\vartheta_{ee} = 4 |U_{e4}|^2 \left( 1 - |U_{e4}|^2 \right)$
of
$\nua{e}$
disappearance,
and
the amplitude
$\sin^22\vartheta_{\mu\mu} = 4 |U_{\mu4}|^2 \left( 1 - |U_{\mu4}|^2 \right)$
of
$\nua{\mu}$
disappearance.

Table~\ref{tab:chi3p1}
shows the comparison of the results of the 3+1 global fit
of short-baseline neutrino oscillation data
without\footnote{
The results without multinucleon interactions in the ``MiniBooNE'' column
in Tab.~\ref{tab:chi3p1}
are different from those in Ref.~\cite{Gariazzo:2015rra}
because we improved the treatment of the MiniBooNE background disappearance
due to neutrino oscillations
according to information kindly given to us by
Bill Louis.
}
and with the multinucleon interactions
in the analysis of MiniBooNE data.
One can see that
in the fit with multinucleon interactions
there is a significant improvement of
the appearance-disappearance parameter goodness-of-fit,
which quantifies the appearance-disappearance tension.
However,
this improvement is not sufficient to solve the problem of the appearance-disappearance tension,
because the value of the appearance-disappearance parameter goodness-of-fit
is still too small.

Figure~\ref{fig:3p1}
shows the comparison of the allowed regions in the
$\sin^22\vartheta_{e\mu}$--$\Delta{m}^2_{41}$
plane without and with the multinucleon interactions
in the analysis of MiniBooNE data.
One can see that the region allowed by appearance data
(inside the blue curve)
is slightly shifted towards larger values of $\Delta{m}^2_{41}$
by taking into account
the multinucleon interactions in the analysis of MiniBooNE data.
However,
the appearance-disappearance tension persists,
since most of the region allowed by appearance data
is excluded by the disappearance data
(the region on the right of the red curve).

\section{Conclusions}
\label{conclusions}

In this paper
we have shown that
taking into account the multinucleon interactions
in the analysis of MiniBooNE data
allows a slightly better fit
of the MiniBooNE low-energy excess
and
induces a shift of the allowed region
in the
$\sin^2 2\vartheta$--$\Delta{m}^2$
plane towards
smaller values of
$\sin^2 2\vartheta$
and
larger values of
$\Delta{m}^2$
in the framework of two-neutrino oscillations.

We performed also a global fit
of short-baseline neutrino oscillation data
in the framework of 3+1 neutrino mixing.
We have shown that taking into account the multinucleon interactions
in the analysis of MiniBooNE data
lead to a decrease of the appearance-disappearance tension.
However,
this effect is not enough in order to
solve the problem of the
appearance-disappearance tension
in the global fit
of short-baseline neutrino oscillation data,
because the value of the appearance-disappearance parameter goodness-of-fit
is still too small.

We conclude that further investigations are needed
for solving the puzzle of the
MiniBooNE low-energy anomaly.
Most notably,
the MicroBooNE experiment at Fermilab
\cite{Szelc:2015dga}
will check if the MiniBooNE low-energy anomaly
is due to photons produced by neutral-current $\nu_{\mu}$ interactions
(for example $\pi^{0}$ production with the detection of only one of the two decay photons).
$\nu_{e}$-induced events
cannot be distinguished from
photon-induced events
in the MiniBooNE mineral-oil Cherenkov detector,
but they can be distinguished in the
MicroBooNE
Liquid Argon Time Projection Chamber.
Eventually,
the SBN experiment at Fermilab
\cite{Antonello:2015lea}
will check in a conclusive way the LSND anomaly
and the neutrino oscillation explanation of
the MiniBooNE data.

\begin{acknowledgments}
We would like to thank Teppei Katori and Bill Louis for information on the MiniBooNE experiment
and
M.B. Barbaro for useful discussions.
M.V.G. would like to thank the Department of Physics of the University of Torino
and the INFN Torino Section
for hospitality and partial support.
The work of M.V.G. was supported by
Deutsche Forschungsgemeinschaft in Sonderforschungsbereich 676.
The work of C.G.
was partially supported by the research grant {\sl Theoretical Astroparticle Physics} number 2012CPPYP7 under the program PRIN 2012 funded by the Ministero dell'Istruzione, Universit\`a e della Ricerca (MIUR).
M.M. acknowledges the support and the framework of the ``Espace de Structure et de r\'eactions Nucl\'eaire Th\'eorique'' (ESNT, \url{http://esnt.cea.fr} ) at CEA.
\end{acknowledgments}

%

\end{document}